\documentclass[10pt,letterpaper]{article}
\usepackage[nottoc]{tocbibind} 
\usepackage[utf8]{inputenc}
\usepackage[T1]{fontenc}
\usepackage[  
  bookmarks=false,
  hyperfootnotes=false,
  hyperindex=false,
  hidelinks,  
  pdftitle={Lara: A Key-Value Algebra underlying Arrays and Relations},  
  pdfauthor={Dylan Hutchison, Bill Howe, Dan Suciu}]{hyperref}
\usepackage{amsmath}
\usepackage{amssymb}
\usepackage{multirow}
\usepackage[left=0.8in,right=0.8in,top=0.5in,bottom=0.8in]{geometry}
\usepackage{pifont}
\usepackage{relsize}

\usepackage{graphicx}
\graphicspath{{.}{./img/}}
\DeclareGraphicsExtensions{.eps,.pdf,.png}

\usepackage{amsthm}
\usepackage{amscd}
\usepackage{ stmaryrd }

\usepackage{dsfont}

\usepackage{accents}

\usepackage{ wasysym }

\usepackage{subfig}
\usepackage{adjustbox}
\usepackage{calc}

\usepackage{todonotes}
\usepackage{tabulary}
\usepackage{array}

\usepackage{varwidth}
\usepackage{verbatim}
\newenvironment{centerverbatim}{%
  \par
  \centering
  \varwidth{\linewidth}%
  \verbatim
}{%
  \endverbatim
  \endvarwidth
  \par
}

\makeatletter
\newcommand\bigincircbin
{%
  \mathpalette\@bigincircbin
}
\newcommand\@bigincircbin[2]
{%
  \mathbin%
  {%
    \ooalign{\hidewidth$#1#2$\hidewidth\crcr$#1\bigcirc$}%
  }%
}
\makeatother
\makeatletter
\newcommand\incircbin
{%
  \mathpalette\@incircbin
}
\newcommand\@incircbin[2]
{%
  \mathbin%
  {%
    \ooalign{\hidewidth$#1#2$\hidewidth\crcr$#1\fullmoon$}%
  }%
}
\makeatother

\def\ojoin{\setbox0=\hbox{$\bowtie$}%
  \rule[-.02ex]{.25em}{.4pt}\llap{\rule[\ht0]{.25em}{.4pt}}}

\def\fullouterjoin{\mathbin{\ojoin\mkern-5.8mu\bowtie\mkern-5.8mu\ojoin}}

\newcommand{\bif}{\text{ if }}
\newcommand{\bthen}{\text{ then }}
\newcommand{\belse}{\text{ else }}

\newcommand{\nul}{\texttt{null}}

\newcommand\uproduct{%
  \mathchoice{\mathbin{\;\rotatebox{90}{$\Bowtie$}}}%
             {\mathbin{\;\rotatebox{90}{$\Bowtie$}}}%
             {\mathbin{\;\rotatebox{90}{\scalebox{0.65}{$\Bowtie$}}}}%
             {\mathbin{\;\rotatebox{90}{\scalebox{0.65}{$\Bowtie$}}}}%
}

\DeclareMathOperator{\join}{\bowtie}
\DeclareMathOperator{\joino}{\bowtie_\otimes}
\DeclareMathOperator{\sjoin}{\hat{\bowtie}}
\DeclareMathOperator{\sjoino}{\hat{\bowtie}_\otimes}
\DeclareMathOperator{\union}{\uproduct}
\DeclareMathOperator{\uniono}{\uproduct_\oplus}

\DeclareMathOperator{\supp}{supp}

\DeclareMathOperator{\map}{map}
\DeclareMathOperator{\mapnz}{mapnz}

\DeclareMathOperator{\promote}{promote}

\DeclareMathOperator{\intro}{intro}
\DeclareMathOperator{\ext}{ext}

\DeclareMathOperator{\flatten}{flatten}
\newcommand{\inv}{^{\raisebox{.2ex}{$\scriptscriptstyle-1$}}}

\usepackage{arydshln}

\usepackage{venndiagram}

\usepackage{url}

\author{Dylan Hutchison, Bill Howe, Dan Suciu \vspace{0.25em} \\  University of Washington}
\title{\textsc{Lara}: A Key-Value Algebra underlying Arrays and Relations}
\date{{\bfseries Draft}: April 1, 2016}
\begin{document}
\maketitle

\begin{abstract}
Data processing systems roughly group into families
such as relational, array, graph, and key-value.
Many data processing tasks exceed the capabilities of any one family,
require data stored across families, or
run faster when partitioned onto multiple families.
Discovering ways to execute computation among multiple available systems,
let alone discovering an optimal execution plan,
is challenging given semantic differences between disparate families of systems.
In this paper we introduce a new algebra, \textsc{Lara}, which underlies and unifies
algebras representing the families above in order to facilitate translation between systems.
We describe the operations and objects of 
\textsc{Lara}---union, join, and ext on associative tables---and show her 
properties and equivalences to other algebras.
Multi-system optimization has a bright future,
in which we proffer \textsc{Lara} for the role of universal connector.
\end{abstract}




\clearpage
\tableofcontents

\clearpage
\addcontentsline{toc}{section}{List of Notation}
\section*{List of Notation}

\begin{table}[h]
\centering 
\begin{tabular}{ c p{15cm} }
\multicolumn{2}{c}{\underline{Components of Associative Tables} (see Section~\ref{sLaraObjects})}\\
$\bar{x}$  & A record, which is a tuple whose components have names called a header \\
$\pi_S(\bar{x})$  & Projection of a record retaining only the components whose name is in set $S$ \\
$\bar{x}.\bar{y}$ & Concatenation of records with disjoint headers \\
$K_A$ & Set of names of key attributes of associative table $A$ \\
$V_A$ & Set of names of value attributes of associative table $A$ \\
$E_S$ & The ``empty table'' with key attributes given by names in set $S$ and no value attributes \\
$\supp(A)$ & The support of an associative table, which is the set of keys that map to a non-default value \\
\multicolumn{2}{c}{}\\
\multicolumn{2}{c}{\underline{Lara Operations on Associative Tables} (see Section~\ref{sLaraOperations})}\\
$\uniono$  & Union \\
$\union$  & Union, when $\oplus$ is unimportant except perhaps that 0 is an identity \\
$\sjoino$  & Strict Join \\ 
$\sjoin$  & Strict Join, when $\otimes$ is unimportant except perhaps that 0 is an annihilator and 1 is an identity \\
$\joino$  & Relaxed Join \\
$\join$  & Relaxed Join, when $\otimes$ is unimportant except perhaps that 0 is an annihilator and 1 is an identity \\
$\ext_f$ & Extension of function $f$ onto an associative table, which may add additional keys \\
$\map_f$ & Special case of $\ext_f$ when $f$ does not add additional keys \\
$\mapnz_f$ & A map that only affects nonzero values \\
$\Pi_V$ & A $\map$ that projects away the value attributes not in $V$ \\
\end{tabular}
\caption{List of Notation}
\label{tNotation}
\end{table}

\clearpage
\section{Introduction}


Data processing systems span several families, including
array, relational, graph, and file systems.
Each family has a unifying logical representation.
Particular systems within a family realize a physical representation 
of its family's logical representation.

For example, the logical data representation of array systems is matrices.
Computation takes the form of linear algebra operations, invoking methods such as 
matrix multiply, reduce, element-wise addition, and matrix sub-reference.
ScaLAPACK is an array system with a physical data representation
following the matrix pattern.  After placing data into the physical format
required by ScaLAPACK, one can make BLAS API calls that perform linear algebra optimizations.
SciDB is another member of the array systems family, with its own physical data format
following the matrix pattern and its own AQL API supporting linear algebra operations.

As another example, the logical data representation of relational systems is relations.
Computation takes the form of relational algebra, invoking methods such as
selection, projection, Cartesian product, union, and aggregation.
PostgreSQL, HStore, Myria, and many other systems are members of the relational family,
each having their own physical data representation following the relation pattern, 
and each having their own API supporting relational algebra calls.

The ``logical algebra / physical API'' pattern holds for graph and file processing as well.
Many graph systems like AllegroGraph and the family of Tinkerpop-compatible databases 
have physical data formats following the graph pattern, 
some in the case of adjacency lists, others as incidence matrices, etc.
The graph systems support APIs of vertex and edge processing with their APIs.
File systems use the logical data representation of files and processing 
in the form of file access (reads and writes, sequential and random).
Particular file systems such as NFS or Lustre have their own physical file formats
implementing the file pattern. They all support the file access operations 
through their APIs in some way.

Isn't it striking that we can perform the same computations 
in systems from each of these system families?
For example, we execute algorithms for matrix inversion 
in array systems, calling the appropriate linear algebra libraries; 
in relational systems, through iterations of joins until convergence; 
in file systems, with the appropriate file read, write, and sort routines;
in MapReduce systems, through passes of maps and reduces.
We can paint similar stories for many other computations
such as convolving images, computing PageRank, training SVMs and other recommenders,
finding shortest paths, constructing data cubes, and plenty more.

Computation in each of these families seems isomorphic, 
in the sense that we can rewrite algorithms written for the API and logical data representation 
of one family into the API and logical representation of the others.
(Of course, computations appearing concise and natural in one family 
may appear convoluted in another.)
Said another way, each family is equivalent in computation expressiveness.

Computation \emph{efficiency}, on the other hand, depends heavily on 
choice of family and even particular systems.
MapReduce systems execute fastest for computations that truly 
fit the pattern of a map and reduce, such as word count.
Other systems tend to execute problems that don't fit the map-reduce pattern
faster, such as array systems for matrix inversion and convolution.
All-to-all shortest paths may execute best on graph systems,
though many relational and array systems are strong contenders.

We therefore have motivation to use a variety of system families 
for execution---not because any one family can express computations 
that the others cannot, but because we gain performance
by leveraging systems from different families for computations
they perform best. 
We call such a hybrid, federated system a polystore system.
The motto of polystore systems is simple:
``Use the right system for the right job.''
Many algorithms can be decomposed into a set of jobs
that execute most efficiently when run on different systems.

A key step to building a polystore 
is to devise a common data representation
which facilitates translation between the languages and systems composing it:
revealing joint optimization opportunities,
promoting more efficient data transfer,
and reducing the semantic gap that programmers face 
when writing code across families.
We propose \emph{associative tables}, 
a new logical data structure that captures core properties
from relations, graphs, spreadsheets, files, and tensors,
all with three operations: 
join ($\join$), union ($\union$), and ext.

In this paper we detail the design of Lara
and connect it to relational and array algebra 
by writing bidirectional translations of their objects and operators.
We highlight the following contributions:
\begin{enumerate}
\item We unify relational, array, and key-value algebras
through the definition of a common data structure and operations.
Translations through Lara facilitate translations between the algebras.
In particular, Lara is a candidate for the proposed BigDAWG polystore system
\cite{elmore2015demonstration}.

We have yet to see whether the ability to translate expressions from one algebra into another
sheds light on what each algebra's operations
mean in the context of another, or in the context of the common Lara algebra.

\item We conject that translating computation through Lara
leads to better performance, gained from running parts of algorithms on different systems.
We conject that the common abstraction of Lara enables 
multi-system optimizations difficult to capture otherwise.
\end{enumerate}

\subsection{Lara's Distinctive Features}
\begin{enumerate}
\item Default values instead of null; tables as total functions. 

Many databases that use the special value null to indicate a non-present or un-stored value 
face a variety of negative issues.
For example, joining tables on nullable attributes requires extra logic,
computing statistics on nullable time series data with nulls has unclear semantics (are nulls included in counts?),
and comparison queries lead to arbitrary choices about whether to order nulls before or after strings or numbers.

Replacing null with a default value solves most issues created by null
because program logic treats default values the same as ordinary values.
In fact, operations on an associative table may modify the table's default value 
the same way operations modify values at rows in the table's support,
whereas users of databases using null would have to remember how their interpretation of null
changes with each successive operation.

The use of default values further enables the interpretation of tables as total functions.
Given any key, as associative table defines a value that the key maps to,
which is the default value if the key is not in the table's support.
Users can query for undefined entries the same way users query for defined entries, 
unlike tables with partial function semantics that return an empty set when queried for a non-present key.


\item Table support need not be minimal.

An associative table's support is the set of keys which map to a non-default value.
Any correct storage scheme must store entries in the table's support.
What about other entries outside the support?

Storage schemes for Lara are free to store any \emph{superset} of a table's support
without affecting correctness. 
This property reduces the burden of functions on associative tables 
because they need not compute the minimal support of their resulting table, 
which may require extra computation.
For example for two associative tables $A$ and $B$ with the same schema, 
the operation $A \union_+ B$ sums the values of $A$ and $B$ into a new table $C$.
It is possible that a key may be in the support of $A$ and $B$ but not $A \union_+ B$,
such as when we add $(-2) + 2 = 0$ and 0 is the default value of the resulting table.
Detecting ``lost support'' resulting from an operation could slow down
otherwise efficient function calls such as the matrix addition operator in the CombBLAS system.
Instead, implementations are free to store spurious ``zero entries''.

\item Open to user-defined types and operators, inheriting their properties when given.


User-defined types and operations pervade real-world applications,
yet many data processing frameworks struggle to support and especially optimize them.
We designed Lara as an open language allowing any user-defined types and operations.
Specifically, we allow custom $\otimes$ in $\sjoino$, $\oplus$ in $\uniono$, and $f$ in $\ext_f$.

Users have the choice of providing as little or as much information about the structure
of their types and operation. 
When given little information, there is little we conclude and instead treat such types and operations as black boxes.
When given rich information about the structure of types and operation, 
such as properties like identities, annihilators, idempotence, and commutativity,
we may conclude more efficient execution strategies
such as those for union in Section~\ref{sUnionExecution}.

Many optimizations derived from known properties for types and functions come from abstract algebra.
Users do have some additional work to identify properties,
but we do not believe the burden of recognizing and annotating properties is unmanageable.
In fact, the approach of annotating functions with properties from abstract algebra 
has already seen industry success in the Summingbird Scala library, which acclaims
``that abstract algebra provides a formal
framework for thinking about and potentially resolving
many thorny issues in distributed processing''
\cite{boykin2014summingbird}.



\item Concise. 
While not a rigorous metric, 
it is pleasing to note that Lara achieves all her expressiveness with only three parameterized operators.
\end{enumerate}

\section{Overview of Lara}
Lara is an algebra over associative tables.
An associative table is a total function from a set of keys to a set of values with finite support.
For example, one could think of a 2-D associative table as a matrix 
with infinitely many rows and columns,
with the condition that only finitely many rows and columns contain a nonzero element.

The union operation $A \uniono B$ ``vertically concatenates'' the values in the union of $A$ and $B$,
grouped by the keys in the intersection of $A$ and $B$, using $\oplus$ to aggregate colliding values.
One can think of union in terms of relational union and aggregate, or in terms of tensor reduce and element-wise sum.

The (strict) join operation $A \sjoino B$ ``horizontally concatenates'' the values in the intersection of $A$ and $B$
by multiplying them with the operation $\otimes$, for values from each pair of entries in the natural relational join of $A$ and $B$'s keys.
One can think of join in terms of tensor product.
A relaxed version of join $A \joino B$ recovers the semantics of relational inner join.

The ext operation $\ext_f A$ behaves like flatmap: to apply function $f$ to each entry of $A$ independently and union together the results.
Ext generalizes relational selection, renaming, and extended projection.
It also has interpretation as ``exploding'' functions like string tokenization.

\subsection{Joint PageRank example}
Figure~\ref{fPageRank} illustrates a Lara algorithm for PageRank \cite{page1999pagerank} 
on the users common to two social networks $S_1$ and $S_2$.
It uses a Datalog-like syntax for naming key and value attributes,
described briefly in Section~\ref{sFurtherWork},
as well as some derived operations from Table~\ref{tNotation}.
The `[0]'s indicate default values. $E$ is the constant empty table.

\begin{figure}[htb]
\begin{align}
&\text{\textbf{Input}: $S_1(src,dst; val [0])$, $S_2(src,dst; val [0])$} \\
&SrcCommon := (S_1(src,\_; 1) \union E(src;)) \sjoin (S_2(src,\_; 1) \union E(src;)) \\
&A := (S_1(src,dst; val) \sjoin SrcCommon(src; val)) \union_{avg} (S_2(src,dst; val) \sjoin SrcCommon(src; val)) \\
&d_{out} := A(src,\_; val) \union_+ E(src;) \\
&d_{out}\inv := \mapnz_{val := val\inv} d_{out}(src;val) \\
&A := A(src,dst; val) \sjoin_* d_{out}\inv(src; val) \\
&r := \mapnz_{val := \text{rand}()}(A(\_, dst; 1) \union E(dst;)) \\
&r := r(dst; val) \sjoin_/ (r(\_; val) \union_+ E(;)) \\
&a := \map_{val := (1-c)} r(dst; val) \sjoin_/ (r(\_; 1) \union_+ E(;)) \\
&\text{For 20 iterations:} \\
&\quad r := (A(src,dst; val) \sjoin_* (\map_{val := val * c} r(dst; val))) \union_+ a(dst; val) \\
&\text{\textbf{Output}: } r 
\end{align}
\caption{PageRank algorithm in Lara}
\label{fPageRank}
\end{figure}

The factor $c$ is a constant such that $1-c$ of the time, the PageRank ``restarts'' from a random page.
Line 2 computes the src nodes common to both $S_1$ and $S_2$ in variable $SrcCommon$.
Line 3 filters $S_1$ and $S_2$ to retain only the edges whose src is in $SrcCommon$,
and stores their union as $A$, running the $avg$ function on edges that appear in both $S_1$ and $S_2$.
Line 4 computes the out-degree of $A$.
Line 5 takes the inverse of the out-degrees, only operating on nonzero values.
Line 6 normalizes $A$ by multiplying each entry in $A$ with the inverse of its out-degree.
Line 7 initializes a random vector $r$ with entries between 0 and 1, on the same support as the $dst$ of $A$.
Line 8 normalizes $r$ by dividing its entries by its sum.
Line 9 constructs a constant vector whose entries equal $(1-c)$ before being normalized, on the same support as $r$.
Line 11 computes the PageRank kernel: $r := A(r * c) + a$.

\section{Lara's Objects: Associative Tables}
\label{sLaraObjects}

An associative table is a total mapping from $m$ key spaces to $n$ value spaces
with finite support, along with a header attaching a name to each key and value space.
By finite support we mean the number of non-default entries of each of the $n$ value spaces is finite.
The following paragraphs define the terms of our definition from the ground up.

We allow the user to supply arbitrary base types like integers and strings.
An $n$-tuple is the Cartesian product of $n$ base types,
as in the 4-tuple $t =$ (3, 9, `abc', 2.5) for which we write $t$'s type as 
$\mathbb{N}\times\mathbb{N}\times\mathbb{S}\times\mathbb{R}$.
Variable names like $t$ may refer either to values or types depending on context.

A record is a tuple with a unique string name for each component, as in
$r =$ (temperature: 73.5, coverage: `low', humidity: 0.75)
for which we write $r$'s type as
(temperature: $\mathbb{R}$, coverage: $\mathbb{S}$, humidity: $\mathbb{R}$).
The set of string names \{temperature, coverage, humidity\} is called $r$'s header,
which we often omit when clear from context.
The order of a record's components does not matter 
since we reference them through names in a header.
A record's dimension is its number of components.
We sometimes refer to components as \emph{attributes} or columns.

We call the operation which removes all components from a record $r$ 
except those components in the header $H$
the projection of $r$ onto $H$ denoted $\pi_H(r)$.
We denote concatenation of records $r_1$ and $r_2$ with disjoint headers as $r_1 . r_2$.
For example, $\pi_{\{\text{humidity}\}} (\text{temperature}: 73.5, \text{coverage}: `low', \text{humidity}: 0.75)
= (\text{humidity}: 0.75)$,
and $(\text{humidity}: 0.75) . (\text{coverage}: `low') = (\text{coverage}: `low', \text{humidity}: 0.75)$.

An associative table $A$
is a total function from a record type $\bar{k}$ which we call the keys of $A$, 
to a record type $\bar{v}$ which we call the values of $A$,
with the requirement that $\bar{k}$ and $\bar{v}$ have disjoint headers
and a distinguished default value $\bar{0}$.
We write $K_A$ for the header of $A$'s keys and $V_A$ for the header of $A$'s values,
so that we may write the disjoint header requirement as $K_A \cap V_A = \emptyset$.

We write $A$ as a table listing mappings from keys $\bar{k}$ to values $\bar{v}$.
We call each mapping an \emph{entry} or \emph{row}.
Keys that do not appear in the table map to $A$'s default value $\bar{0}$.
We write the type of $A$ as
\[ A : [[ \bar{k} \mapsto \bar{v} : \bar{0} ]] \]
We call $A$ as a function via the expression $A(\bar{k})$
which yields the value associated with $\bar{k}$, 
usually dropping nested parentheses for readability.
Associative tables have key attributes, value attributes,
a key dimension $m$, and a value dimension $n$, 
corresponding to the parts from the table's keys and values.
The dimension of an associative table is its key dimension $m$,
which tells us the number of items we must pass to $A$ as a function
in order to uniquely identify $n$ values under $A$'s image.

The support of $A$, written $\supp(A)$,
is the set of keys that map to a non-default record under $A$.
One way to write the support is
$\supp(A) = A\inv ( V \setminus \{\bar{0}\} )$,
where $V$ is the set of all value records 
and $\bar{0}$ is the default value record.

\begin{figure}
\centering
\subfloat[`Part' table $P$]{ 
  \begin{tabular}{c|cc}
   & [white] & [0] \\
  pid & color & wgt \\
  \hline
  p01 & blue & 3 \\
  p02 & red & 4 \\
  p04 & blue & 2 \\
  \end{tabular}
  \label{tPart}
}
\subfloat[`Supplier' table $S$]{ 
  \begin{tabular}{c|cc}
   & [unknown] & [WA] \\
  sid & fav & state \\
  \hline  
  s01 & blue & WA \\
  s02 & red & NJ \\
  s04 & blue & NJ \\
  \end{tabular}
  \label{tSupplier}
}
\subfloat[`Request' table $R$]{ 
  \begin{tabular}{cc|cc}
   & & [0] & [n] \\
  sid & pid & qty & urgent \\
  \hline  
  s01 & p02 & 3 & n \\
  s02 & p03 & 1 & n \\
  \end{tabular}
  \label{tRequest}
}
\caption{`Part-Supplier-Request' database}
\label{fPartSupplierRequest}
\end{figure}

See Figure~\ref{fPartSupplierRequest} for example associative tables.
In the figure's presentation, the labels immediately above the horizontal bar
are key/value names and the listings below them are rows of mappings.
The attributes to the left of the vertical bar are keys, 
and the attributes to the right are values.
Value attributes have a default value indicated in brackets above the attribute name.

From the tables in Figure~\ref{fPartSupplierRequest}, 
we see that $P(\text{pid}: p02) = (\text{color}: red, \text{wgt}:4)$.
The projection $\pi_{\text{wgt}} P(\text{pid}: p02) = (\text{wgt}:4)$.
Keys not explicitly listed in the table 
return the default value for each attribute, e.g.
$P(p03) = (white, 0)$.


Zero-dimensional associative tables have no key attributes.
Such an associative table $A$ acts as a function from \emph{unit}, 
the set of 0-tuples, to its value attribute type.
It has a single mapping from the 0-tuple $()$ to a record value $\bar{v}$
which by convention coincides with the table's default value $\bar{0}$.

Tables may also have zero value dimension, where every key maps to the 0-tuple $()$.
We call tables with no values \emph{empty tables}, 
because these tables necessarily have zero support 
since every key must map to the value $()$
which must be the default value as it is the only value of the unit type.
The notation $E_K$ denotes the empty table with key attributes $K$.
A table with neither keys nor values, $E$, has the one mapping $() \mapsto ()$.

\subsection{Translating Lara's and Other Algebras' Objects}

In this section we show mappings  
between associative tables and the objects of other algebras.
We defer mapping the operations of the Lara algebra 
to and from operations of other algebras to Section~\ref{sTranslateOperations}.

Most mappings between associative tables and other objects are not unique,
in the sense that an associative table could be represented by many other 
objects, or an object could be represented by many associative tables.
We eliminate ambiguity in these cases by grouping together 
all objects that an associative table could map to (and/or vice versa)
by an \emph{equivalence class}, so that we can claim isomorphism 
between equivalence classes of associative tables and equivalence classes of the other objects.

We already know one equivalence class between 
pictures of associative tables we write down, as in Figure~\ref{fPartSupplierRequest}:
$A$ and $B$ are equivalent if they have the same headers and support,
meaning that $A$ and $B$ differ only in the listing of default values.
For example, appending the entry ``$(\text{p76}) \mapsto (\text{white}, 0)$''
to the `Parts' table does not change the table at all, because
the entry already exists by the semantics of default values.

More subtly, associative table pictures with the same entries 
but written in different row orders or column orders 
(so long as key columns do not mix with value columns) are equivalent
In other words, associative tables are invariant 
to permuting the rows or columns of their picture.

Implementations pick the ``most efficient'' 
member of an equivalence class as its representative. 
In the case of a class of associative table pictures,
this is the picture that does not include any entries 
with all default values, since writing these is superfluous.
We also tend to write rows in an order given by an ordering on their keys.
The order of columns is mostly arbitrary.

Real data structures might place more meaning on the order of rows and columns
since they may correspond to a physical ordering of data on disk.
Similarly, real data structures may materialize an entry with all default values for some reason.

We now list mappings between associative tables and five objects
illustrated in Figure~\ref{fRep}:





\newlength{\mycolsep}
\setlength{\mycolsep}{3pt}
\newlength\myheight
\newlength\myheighttmp
\setlength{\myheight}{150pt}

\begingroup
\setlength{\tabcolsep}{\mycolsep}
\renewcommand{\arraystretch}{1.2} 
\begin{figure}[htb]
\centering
\subfloat[CSV file]{
\begin{adjustbox}{raise=39pt} 
\begin{tabular}{l}
r,c,v\\
0730,Alice,30 \\
0730,Casey,30 \\ 
1145,Bob,60   \\
1145,Joe,60   \\
1400,Bob,15   \\
1400,Casey,15 
\end{tabular}
\end{adjustbox}
}%
\hfil
\subfloat[Sparse matrix]{
\begin{adjustbox}{} 
\shortstack{
  \begin{tabular}{c|ccc}
  r & 0730 & 1145 & 1400 \\
  \hline
  i & 1 & 2 & 3  \\
  \end{tabular} \\
  \begin{tabular}{c|cccc}
  c & Alice & Bob & Casey & Joe \\
  \hline
  j & 1 & 2 & 3 & 4 \\
  \end{tabular} \\
  \begin{tabular}{c|ccc}
  v & 15 & 30 & 60 \\
  \hline
  k & 1 & 2 & 3 \\
  \end{tabular} \\
  $\begin{bmatrix}
  2 & & 2 & \\
  & 3 & & 3 \\
  & 1 & 1 & \\
  \end{bmatrix}$
}%
\end{adjustbox}
}
\hfil
\shortstack{
\subfloat[Relation]{
  \begin{tabular}{ccccc}
  id & Alice & Bob & Casey & Joe \\
  \hline  
  0730 & 30 & \nul & 30 & \nul \\
  1145 & \nul & 60 & \nul & 60 \\
  1400 & \nul & 15 & 15 & \nul \\
  \end{tabular}
} \\
\subfloat[Spreadsheet]{
\begin{adjustbox}{raise=8pt} 
  \begin{tabular}{l|cccc}
   & Alice & Bob & Casey & Joe \\
  \hline  
  0730 & 30 & & 30 & \\
  1145 & & 60 & & 60 \\
  1400 & & 15 & 15 & \\
  \end{tabular}
\end{adjustbox}
}%
}
\hfil
\subfloat[Graph adjacency matrix]{
\begin{adjustbox}{raise=-17pt} 
\begin{tikzpicture}
\node (v1) at (0,1.5) {0730};
\node (v4) at (0,0) {1145};
\node (v7) at (0,-1.5) {1400};

\node (v2) at (3,2.25) {Alice};
\node (v5) at (3,0.75) {Bob};
\node (v3) at (3,-0.75) {Casey};
\node (v6) at (3,-2.25) {Joe};

\draw  (v1) -- (v2) node[midway, sloped, above] {30};
\draw  (v1) -- (v3) node[near start, sloped, above] {30};
\draw  (v4) -- (v5) node[near end, sloped, above] {60};
\draw  (v6) -- (v4) node[near start, sloped, above] {60};
\draw  (v7) -- (v3) node[near end, sloped, above] {15};
\draw  (v7) -- (v5) node[midway, sloped, above] {15};
\end{tikzpicture}
\end{adjustbox}
}
\caption{Five representations of an associative table}
\label{fRep}
\end{figure}
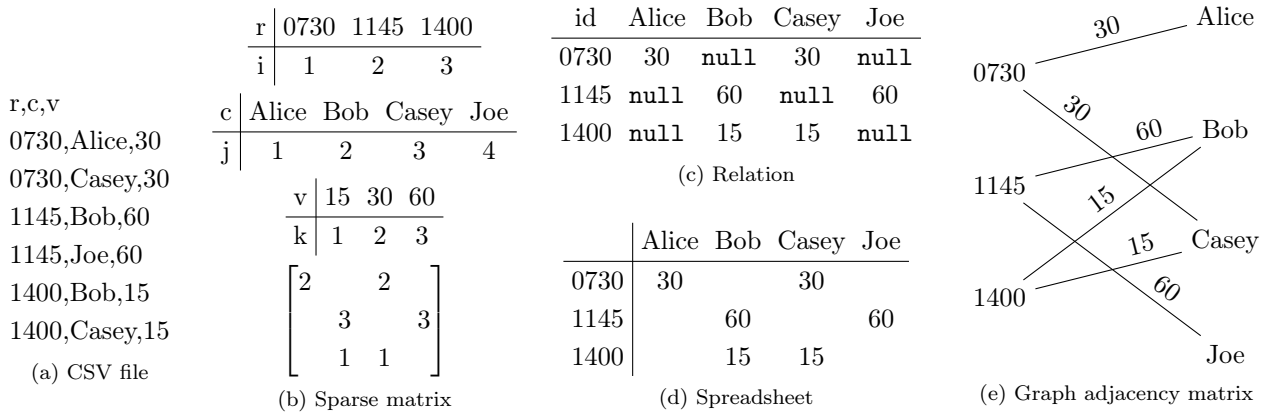
\endgroup

\begin{enumerate}
\item 
A \emph{delimited file} contains text data
stored between occurrences of column separator and row separator characters.
For example, a CSV file delimits columns by commas and rows by line feeds or carriage returns.
TSV files use tab characters in place of commas.
Both file types may include a header on its first line with names for each column.

We can write an associative table $A$ with key dimension $m$ and value dimension $n$
as a delimited file by writing rows of $m+n$ columns each.
First we fix an order of key attributes and value attributes.
We then write one delimited row for the header of keys and values.
Then for each key $\bar{k} \in \supp(A)$, 
we write a row with $\bar{k}$ and $A(\bar{k})$.
There is no need to write rows with all default values.

The reverse translation from a delimited file to an associative table is more ambiguous
because (1) there is no distinction between keys and values and (2)
rows may vary in number of columns. 
Delimited files may also list the same key multiple times.
One way to handle duplicate keys is to provide a ``collision function'' $\oplus$
such that the associative table formed by reading the delimited file stores the value 
corresponding to the sum of values having the same duplicate key.


\item 
A \emph{relation}, or relational database table, 
is a set of records with the same record type.
We call the record type the relation's schema.

We can write an associative table $A$ as a relation $R$
by fixing the relation's schema as the concatenation of 
$A$'s key and value record types, and defining the
contents of the relation as records containing the key and corresponding value
for each key in $A$'s support.
The attributes of $R$ corresponding to the keys of $A$
form a compound key for $R$ by construction,
or in other words, a set of keys uniquely identifies a row in $R$.
Databases typically mark the compound key as the primary key,
meaning that it is the set of attributes used for 
database access and for joining tables.

To form the reverse translation from a relation $R$ to an associative table $A$, 
identify a subset of $R$'s attributes that form a compound key, 
and let those attributes be the key attributes of $A$.
A compound key always exists for $R$, because in the worst case 
the set of all $R$'s attributes form a compound key.
There may be more than one choice of compound key.
As in the case of delimited files, we may even choose key attributes for $A$
that are not a compound key for $R$, instead defining a collision function $\oplus$ 
which sums together values having the same key.

The special value \texttt{null} may take the place of a default value.
This is fine so long as implementations track the meaning of \texttt{null} 
as a default value.

\item 
A \emph{key-value store} such as a hash table or many NoSQL databases
has a clear mapping to and from associative tables:
the mappings from keys to values of one are identical in the other.
Spreadsheets are examples of two-dimensional key-value stores.

\item 
An \emph{associative array}~\cite{kepner2016math}
is a data structure mapping $m$ key spaces $K_1, \dots, K_m$ to a value space $V$
with the condition that $V$ has a semi-ring structure $(V,0,1,\oplus,\otimes)$
and that the associative array has finite support---that a finite number of keys map to 0.

The differences between $m$-dimensional associative tables and associative arrays
are (1) that the tables have a name for each dimension (defined in the header),
(2) that the tables store $n$ values at each entry instead of one,
and (3) that there is no pervasive semi-ring structure requirement.
Requirements on identities and annihilators come into play when considering operations.

Associative arrays have known translations between spreadsheets, 
(relational, key-value, and array) databases,
and graphs \cite{kepner2015unified}.
Because associative tables have translations to and from associative arrays,
we see that associative tables also have translations 
with spreadsheets, databases and graphs transitively, and 
constructing direct translations is one step away.

\item 
A \emph{tensor} of $m$ dimensions
maps $m$ natural number keys to a space of values $V$.
Sparse tensors additionally have a distinguished zero value $0 \in V$ 
that is the default value for entries not explicitly represented.

We can construct a sparse tensor representing an associative table $A$
by letting $V$ be $A$'s value record type.
We can transform $A$'s key record type into a tuple of integers by
defining an order on $A$'s header and 
defining lookup tables that assign an integer to each key in the support of $A$.
These auxiliary lookup tables have finite size by the definition of an associative table.

To form an associative table from a sparse tensor, 
construct a header by assigning names `1', `2', ..., `$m$'
to the tensor's dimensions, and the remaining construction follows.

\item The adjacency tensor of a \emph{hyper-graph}, 
with hyper-edges between $m$ nodes having $n$ labels each.
Given a graph's adjacency tensor, the translation to and from associative tables
is identical to that of tensors.

\end{enumerate}

Because all the above data structures are equivalent up to translation,
each object's storage scheme is applicable to associative tables,
which allows the privilege of choosing an optimal storage scheme relative to a cost model.
For example, database storage schemes often range between 
row store and column store strategies for the case 
of more than one value attribute.

Suppose an associative table has $m$ keys and $n$ values.
A row store stores $n$ values together, sorted and indexed by the $m$ keys.
A column store stores each of the $n$ values separately,
each sorted and indexed by the $m$ keys.
We realize a row storage strategy via a single associative table with $n$ values,
whereas we realize a column storage strategy via $n$ associative tables with 1 value each.
Hybrid storage strategies grouping some values together
(in groups often called column families or locality groups)
while storing each group separate.

\section{Lara's Operations: Union, Join, Ext}
\label{sLaraOperations}

We define three operations on associative tables---join, union, and ext---%
parameterized by user-definable ``sum'', ``multiply'', and ``flatmap'' functions respectively.



\subsection{Table Union: Vertically Concatenate and Sum}
\label{sUnion}
Union is a binary operation on tables, written as $A \uniono B$ and
parameterized by a binary operator $\oplus$ over scalars.
We omit the $\oplus$ subscript when insignificant.
Informally, union ``aggregates'' $A$ and $B$
onto their common key attributes, using $\oplus$ as a collision function,
and ``vertically concatenates'' their contents, again executing $\oplus$ on collisions.

Suppose $A$ and $B$ have types (attribute types and default values omitted)
\begin{align*}
&A : [[a_1, \dots, a_m, c_1, \dots, c_n \mapsto x_1, \dots, x_q, z_1, \dots, z_r]] \\
&B : [[c_1, \dots, c_n, b_1, \dots, b_p \mapsto z_1, \dots, z_r, y_1, \dots, y_s]] 
\end{align*}
(the $a$'s and $x$'s are unique to $A$, the $b$'s and $y$'s unique to $B$, 
and the $c$'s and $z$'s common to both $A$ and $B$)
and we have a binary operation $\oplus$ applicable to each value attribute's type,
and for which every value attribute's default value 0 is an additive identity.
Then we define 
\[ A \uniono B : [[c_1, \dots, c_n \mapsto x_1, \dots, x_q, z_1, \dots, z_r, y_1, \dots, y_s]] \]
which has key attributes equal to the intersection of $A$ and $B$'s key attributes,
and value attributes equal to the union of $A$ and $B$'s value attributes, as
\begin{align*}
(A \union_\oplus B)(c_1, \dots, c_n) := 
  \Big( 
  &v_x: \bigoplus_{a_1, \dots, a_m} \pi_x A(a_1, \dots, a_m, c_1, \dots, c_n), \text{ (for each $x$)}  \\
  &v_z: \bigoplus_{a_1, \dots, a_m} \pi_z A(a_1, \dots, a_m, c_1, \dots, c_n) \oplus \bigoplus_{b_1, \dots, b_p} \pi_z B(c_1, \dots, c_n, b_1, \dots, b_p), \text{ (for each $z$)} \\
  &v_y: \bigoplus_{b_1, \dots, b_p} \pi_y B(c_1, \dots, c_n, b_1, \dots, b_p) \text{ (for each $y$)}
  \Big)
\end{align*}

This definition assumes the same $\oplus$ operation applies to every value attribute.
In practice we often wish to apply a different $\oplus$ operation on different value attributes.
For example, we may wish to concatenate a value attribute consisting of strings
at the same time we sum a value attribute consisting of integers.

We therefore extend the $\union$ operator to take a tuple of $\oplus$ operators
$(\oplus_{x_1},\dots,\oplus_{x_q}, \oplus_{z_1},\dots,\oplus_{z_r},\oplus_{y_1},\dots,\oplus_{y_s})$
that each act on an individual value attribute.
This extension does not increase expressiveness,
because as shown in Section~\ref{sDecompose},
associative tables' value attributes 
can be losslessly decomposed and recomposed. 

In relational systems, 
$A \uniono B \equiv \gamma_{K_A \cap K_B, \oplus(V_A \cup V_B)}(A \cup B)$
where $K_A$ are the key attribute names of A 
and $V_A$ are the value attribute names of A.
The associative table union operator 
is a relational union ($\cup$) and an aggregation ($\gamma$),
over all value attributes $V_A \cup V_B$ with operation $\oplus$,
grouped by the key attributes common to both tables $K_A \cap K_B$.

In array systems,
$A \uniono B$ are `reduce' operations on $A$ and $B$
followed by an element-wise sum.

\subsubsection{Execution of Union}
\label{sUnionExecution}
The union operation implements a form of processing 
known as \emph{structural recursion}:
to recursively apply a function to $A$'s contents
and combine results via another function.
Expanding on known theory behind structural recursion
\cite{buneman1995principles},
we can optimize the execution of union 
if we known additional properties of $\oplus$.

We show optimizations to union on a call to $h(S)$ in Table~\ref{tPlusProp},
where $h$ is a structurally recursive function over a set $S$ 
containing the values from a value attribute,
grouped by key attributes, under aggregation during $\uniono$.
In other words, execution of $\uniono$ involves
running $h(S)$ on every group of values, 
grouped by the key attributes in the result of $\uniono$,
reducing those groups to a single value each.

The top row of Table~\ref{tPlusProp} shows an execution strategy for union
with the minimum amount of known properties of $\oplus$ that guarantee correctness,
which are that the default value 0 of the value attribute under $\oplus$ 
must be an identity for $\oplus$.
Subsequent rows of Table~\ref{tPlusProp} show 
how knowing additional properties of $\oplus$ lead to more efficient execution strategies.

\begin{table}[htb]
\centering
\begin{tabular}{ll}
Properties of $\oplus$ & Effect on $\uniono$'s structural recursion \\
\hline
\begin{tabular}{l}
Identity 0 
\end{tabular} & {
    \begin{tabular}{l}
    Requires total order $<$ on keys being aggregated.
    Linear execution: \\
    $\quad h(\emptyset) = 0$ \\
    $\quad h(x \sswarrow S) = e_1(x) \oplus h(S)$ \\
    where $x \sswarrow S$ reads as ``remove least element $x$ from $S$''.
    \end{tabular}} \\
\hline
\begin{tabular}{l}
Associative \\ 
Identity 0  
\end{tabular} 
& {
    \begin{tabular}{l}
    Requires total order $<$ on keys being aggregated.
    Parallel execution: \\
    $\quad h(\emptyset) = 0$ \\
    $\quad h(\{x\}) = e_1(x)$ \\
    $\quad h(A \sqcup B) = h(A) \oplus h(B)$ \\
    where $\sqcup$ is disjoint union ($A \cap B = \emptyset$).
    \end{tabular}} \\
\hline
\begin{tabular}{l}
Associative \\
Identity 0 \\
Commutative 
\end{tabular} & {
    \begin{tabular}{l}
    No order requirement. \\
    Parallel execution, as above. 
    \end{tabular}} \\
\hline
\begin{tabular}{l}
Associative \\
Identity 0 \\
Commutative \\
Idempotent \\
 $\quad$ ($a \oplus a = a$)
\end{tabular} & {
    \begin{tabular}{l}
    No order requirement. Parallel execution: \\
    $\quad h(\emptyset) = 0$ \\
    $\quad h(\{x\}) = e_1(x)$ \\
    $\quad h(A \cup B) = h(A) \oplus h(B)$ \\
    where $\cup$ is union allowing overlap ($A \cap B \neq \emptyset$ is allowed).
    \end{tabular}} \\
\end{tabular}
\caption{Implementation effects of additional $\oplus$ properties}
\label{tPlusProp}
\end{table}

\subsection{Strict Table Join: Horizontally Concatenate and Multiply}
Join is a binary operation on tables
parameterized by a binary operator $\otimes$ over scalars.
Join takes the ``natural join'' of $A$ and $B$'s key attributes
using $\otimes$ to combine matching value attributes,
which has the effect of ``horizontally concatenating'' $A$ and $B$'s contents.

We write the strict version of join as $\sjoino$,
deferring a relaxed version of join to Section~\ref{sRelaxedJoin}.
We omit the $\otimes$ subscript when insignificant.
The strict version has two extra constraints
that guarantee an $\otimes$ multiplication always runs on 
a scalar from $A$ and a scalar from $B$.
These requirements are that
\begin{enumerate}
\item no key attribute of $A$ is a value attribute of $B$ and vice versa 
($K_A \cap V_B = K_B \cap V_A = \emptyset$), and
\item $A \sjoino B$ has value attributes equal to the intersection 
of $A$ and $B$'s value attributes
($V_{A \sjoino B} = V_A \cap V_B$).
\end{enumerate}

Suppose $A$ and $B$ have types (attribute types and default values omitted)
\begin{align*}
&A : [[a_1, \dots, a_m, c_1, \dots, c_n \mapsto x_1, \dots, x_q, z_1, \dots, z_r]] \\
&B : [[c_1, \dots, c_n, b_1, \dots, b_p \mapsto z_1, \dots, z_r, y_1, \dots, y_s]] 
\end{align*}
and we have a binary operation $\otimes$ applicable to the type of each common value attribute
$z_1, \dots, z_r$,
and for which $z_1, \dots, z_r$'s default values are annihilators of $\otimes$
resulting in the scalar defined by the left default value $\otimes$ the right default value.
Then we define 
\[ A \sjoino B : [[a_1, \dots, a_m, c_1, \dots, c_n, b_1, \dots, b_p \mapsto z_1, \dots, z_r]] \]
which has key attributes equal to the union of $A$ and $B$'s key attributes,
and value attributes equal to the intersection of $A$ and $B$'s value attributes, as
\begin{align*}
(A \sjoino B)&(a_1, \dots, a_m, c_1, \dots, c_n, b_1, \dots, b_p) := \\
  &\Big(
  v_z: \pi_z A(a_1, \dots, a_m, c_1, \dots, c_n) \otimes \pi_z B(c_1, \dots, c_n, b_1, \dots, b_p), \text{ (for each $z$)} 
  \Big)
\end{align*}
The requirement that the default values of $z_1, \dots, z_r$ be annihilators of $\otimes$ may be relaxed.
See Appendix~\ref{aJoinZeros} for details.

As with $\union$, the definition of $\sjoin$ assumes the same $\otimes$ operation applies to every value attribute.
We may extend the $\sjoin$ operator to take a tuple of $\otimes$ operators
$(\otimes_{z_1},\dots,\otimes_{z_r})$
that acts on individual value attributes.
Allowing multiple $\otimes$ operations in a join does not increase expressiveness,
because as shown in Section~\ref{sDecompose},
associative tables' value attributes 
can be losslessly decomposed and recomposed,
so that we can simulate a join of $n$ $\otimes$ operations
via the union of $n$ joins of one $\otimes$ operation each.

In relational systems, 
$A \sjoino B \equiv \pi_{(V \otimes V' \text{ as } V)}(\pi_{K_A,V}(A) \bowtie \rho_{V \to V'}(\pi_{K_B,V}(B)))$
where $V$ is the set of value attributes common to $A$ and $B$
and $K_A$ is the set of key attributes of $A$.
The strict associative table join operation
is a relational natural join on the key attributes of $A$ and $B$
after projecting away value attributes not common to both $A$ and $B$,
and then using an extended projection to multiply 
the common value attributes and retain the original names.
When $A$ and $B$ have disjoint key attributes, 
the natural join is a Cartesian product.

In array systems, 
$A \sjoino B$ is a tensor product.

\subsection{Ext: Flatmap}

\begin{figure}[htb]
\centering
\begin{minipage}[c]{.5\textwidth}
\centering
\subfloat[$D$]{ 
  \begin{tabular}{c|c}
   & [`'] \\
  doc & txt \\
  \hline
  d01 & she sells seashells \\
  d02 & shells she sells are shells from sea \\
  d04 & so she sells seashore shells \\
  \end{tabular}
  \label{fExtD}
} \\ 
\subfloat[$\ext_{f_1}(D)$]{ 
  \begin{tabular}{c|c}
   & [0] \\
  doc & cnt \\
  \hline
  d01 & 3 \\
  d02 & 7 \\
  d04 & 5 \\
  \end{tabular}
  \label{fExtD1}
} 
\end{minipage}
$\quad$
\subfloat[$\ext_{f_2}(D)$]{ 
  \begin{tabular}{cc|c}
   & & [0] \\
  doc & wrd & cnt \\
  \hline
  d01 & she & 1 \\
  d01 & sells & 1 \\
  d01 & seashells & 1 \\
  d02 & shells & 2 \\
  d02 & she & 1 \\
  d02 & sells & 1 \\
  d02 & are & 1 \\
  d02 & from & 1 \\
  d02 & sea & 1 \\
  d04 & so & 1 \\
  d04 & she & 1 \\
  d04 & sells & 1 \\
  d04 & seashore & 1 \\
  d04 & shells & 1 \\
  \end{tabular}
  \label{fExtD2}
}
\caption{Example ext operations. $f_1(doc,txt)=\text{wordcount}(txt)$; $f_2(doc,txt)=\text{tokenize}(txt)$}
\label{fExt}
\end{figure}

Buneman et al coined the operation `ext' to represent 
the extension of a function $f: t \to \{s\}$ on a collection's elements
to structural recursion over the whole collection $\ext(f): \{t\} \to \{s\}$
\cite{buneman1995principles}.
One way to define ext is by the `flatmap' $\ext := \flatten \circ \map$,
where $\flatten: \{\{t\}\} \to \{t\}$ unions a set's elements, and
when we have a function $g: t \to s$, we define $\map(g): \{t\} \to \{s\}$
as the application of $g$ to a set's elements.

Intuitively, we see from the flatmap composition 
that ext applies a function $f$ independently to each element in a set,
and then unions (``flattens'') the results together.
Ext behaves like `map' when $f$ returns singleton sets, 
like a `filtering map' when $f$ returns singleton or empty sets,
and like an `explode' operation\footnote{
  An early term used for an operation that split 
  a string of text into rows or columns containing its words
  is `BREAK' \cite{stonebraker1983document}.
  We use the term `explode' in the same sense.
}
when $f$ returns sets with more than one element.
The same intuition carries to our upcoming definition of ext on associative tables.

We include an ext operation in Lara to add map, filter, explode, and rename 
capabilities to associative tables all with one function.
Suppose an associative table $A$ has type
\[ A : [[ a_1,\dots,a_m \mapsto x_1,\dots,x_n : 0_1,\dots,0_n ]] \]
and we have a function
\[ f : a_1\times\dots\times a_m\times x_1\times\dots\times x_n \to  (b_1\times\dots\times b_{m'} \to y_1\times\dots\times y_{n'}) \]
which is a function on $A$'s keys and values 
that returns a function from new keys $b_1,\dots,b_{m'}$ to new values $y_1,\dots,y_{n'}$.

Further suppose that $f$ obeys the restrictions
\begin{enumerate}
\item $\forall  a_1,\dots,a_m; f(a_1,\dots,a_m,0_1,\dots,0_n) = \text{ constants } (0'_1,\dots,0'_{n'})$ and
\item $\forall  a_1,\dots,a_m,x_1,\dots,x_n; f(a_1,\dots,a_m,x_1,\dots,x_n)$ has finite support.
\end{enumerate}
The first requirement forces $f$ to return new default values $0'_1,\dots,0'_{n'}$ 
whenever all old default values $0_1,\dots,0_n$ are passed for $x_1,\dots,x_n$.
The finite support requirement forces $f$ to return values other than $0'_1,\dots,0'_{n'}$
only finitely many times,
enabling the enumeration of mappings to non-default values in tabular form.

The two restrictions ensure we may correctly extend 
$f$, a function from keys and values to additional keys and new values,
to $\ext_f$, a function on associative tables:
\begin{align*}
\ext_f(A) &: [[a_1,\dots,a_m,b_1,\dots,b_{m'} \mapsto y_1,\dots,y_{n'} : 0'_1,\dots,0'_{n'} ]] \\
\ext_f(A)(a_1,\dots,a_m,b_1,\dots,b_{m'}) &:= f((a_1,\dots,a_m).A(a_1,\dots,a_m)) (b_1,\dots,b_{m'})
\end{align*}

The flatmap intuition of $\ext_f(A)$ is that
ext applies a function from a row of an associative table to a new associative table (with fixed type)
to each row or $A$, and then takes the union of (``flattens'') all the independently generated associative tables.
The finite support requirement ensures the resulting associative table is well-formed.

We recover the behavior of map from $\ext_f$ when $f$ has no $b$s and a single $y$ in its type.
A filtering map is $\ext_f(A)$ when $f$ maps some non-default old values to default new values,
effectively shrinking $A$'s support.
The explode case occurs when $f$ has more than one $b$.
Ext behaves as a rename operation on values when $f$ has different names in the value header of its type
and returns singleton sets of records containing the same elements in its definition.
We rename key names by an ext which adds new key values with the same contents as an old key value,
followed by a union that removes the old key values 
(no aggregation occurs in the union because the support remains constant).

Figure~\ref{fExtD1} shows an example of a mapping $\ext_f$ where $f$ is the wordcount function, 
and Figure~\ref{fExtD2} shows an exploding $\ext_f$ where $f$ is the string tokenizing function.
Wordcount maps a string to the number of words inside it.
Tokenize maps a string to a table, which maps a word to the number of occurrences of that word in the string.
The default value of the new tables are tokenize($\cdot$, `') = 0.

The map, filter, and rename cases of ext have clear analogues 
as the extended projection, selection, and rename operators in relational systems
and the apply and array subset operators in array systems.
The full ext function, however, has more power 
than expressions we could write with simple combinations of relational and array operators.
Some relational and array systems include ext-like extensions such as 
the \texttt{string\_to\_array} function in Postgres or the \texttt{FULLTEXT} index in MySQL, 
the val2col explosion function in D4M \cite{kepner2012dynamic},
and the one-to-many mapping operator in the data cleaning / entity resolution language in \cite{galhardas2001declarative}.



\subsection{Relaxed Table Join}
\label{sRelaxedJoin}
We now present a relaxed and more usable version of join derived from strict join.
The relaxed version mirrors relational inner join on the tables' key attributes,
with matching value attributes multiplied as in strict join and non-matching value attributes retained.
We write relaxed join as $A \joino B$ 
(removing the accent from $A \sjoino B$).

\subsubsection{Automatic Multiply-by-one on Missing Value Attributes}
Suppose $B$ has a value attribute $`Bval'$ not present in $A$.

Prior to the join operation, if $B$ has value attributes not present in $A$,
we introduce those value attributes to $A$ using  
$\ext_{supone(v)}(A)$ for each such value attribute $v \in V_B \setminus V_A$.
Given that $A$'s previous value attributes are $V_A$,
the function $supone(v) : K_A \times V_A \to V_A . v$
returns the same value attributes, plus a new value attribute $v$
with value 1 at least one previous value attribute is non-default 
and 0 if all previous value attributes are default.
where $f$ is the constant function that always returns 1,
for each $v \in V_B \setminus V_A$, where $V_A$ is the set of value attributes of $A$.
We similarly introduce value attributes present in $A$ but not $B$ to $B$.

Put another way, the effect of the $\ext_{supone}$ operations
is to add a new value attribute to $A$ for each value attribute unique to $B$
with value 1 in the rows in the support of $A$ and default value 0, and vice versa.
After adding the missing value attributes, we join the tables as in $A \sjoino B$.
The `1' and `0' from $\ext_{supone}$ are placeholders 
for the identity and annihilator of the $\otimes$ in $A \joino B$.

The overall effect is that $A \joino B$ has the union of $A$ and $B$'s value attributes 
instead of the intersection.
The values of value attributes unique to $A$ and in the support of $A$ 
are multiplied by 1.
The effect mirrors the semantics of a relational inner join operation,
including a domain caveat best illustrated by example.

\begin{figure}
\centering
\subfloat[$P$]{ 
  \begin{tabular}{cc|c}
   & & [white]  \\
  cid & pid & color \\
  \hline
  M & p01 & blue \\
  T & p01 & red  \\
  M & p02 & green  \\
  W & p01 & yellow \\
  \end{tabular}
  \label{fInnerJoinP}
} $\quad$
\subfloat[$\ext_{supone(`state')}(P)$]{ 
  \begin{tabular}{cc|cc}
   & & [white] & [0] \\
  cid & pid & color & state \\
  \hline
  M & p01 & blue & 1 \\
  T & p01 & red & 1  \\
  M & p02 & green & 1 \\
  W & p01 & yellow & 1 \\
  \end{tabular}
  \label{fInnerJoinPZ}
} $\quad$
\subfloat[$S$]{ 
  \begin{tabular}{cc|c}
   & & [GA] \\
  cid & sid & state \\
  \hline  
  M & s01 & WA \\
  M & s02 & NJ \\
  T & s02 & DE \\
  F & s01 & CA \\
  \end{tabular}
  \label{fInnerJoinS}
} $\quad$
\subfloat[$\ext_{supone(`color')}(S)$]{ 
  \begin{tabular}{cc|cc}
   & & [GA] & [0] \\
  cid & sid & state & color \\
  \hline  
  M & s01 & WA & 1 \\
  M & s02 & NJ & 1 \\
  T & s02 & DE & 1 \\
  F & s01 & CA & 1 \\
  \end{tabular}
  \label{fInnerJoinSZ}
}

\subfloat[$P \joino S =$\newline $ \ext_{supone(`state')}(P) \sjoino \ext_{supone(`color')}(S)$]{ 
  \begin{tabular}{ccc|cc}
   & & & [white] & [GA] \\
  cid & pid & sid & color & state \\
  \hline
  M & p01 & s01 & blue & WA \\
  M & p01 & s02 & blue & NJ \\
  M & p02 & s01 & green & WA \\
  M & p02 & s02 & green & NJ \\
  T & p01 & s02 & red & DE  \\
  \end{tabular}
  \label{fInnerJoinPZSZ}
}
\caption{Automatic multiply-by-one in $P \joino S$, mirroring relational inner join}
\label{fInnerJoin}
\end{figure}

Figure~\ref{fInnerJoin} shows a relaxed join $P \joino S$.
The join result is exactly what we would expect from a relational join,
but unlike strict table join, 
we must take care to interpret $P \joino S$ correctly.
In table $P$, we see that $P(T, p01) = red$.
The entry $(P \joino S)(T,p01,s02)$ correctly reflects this fact
by listing $red$ under the $color$ value attribute.
However, the entry $(P \joino S)(T,p01,s01)$
evaluates to the color $white$.
In other words, the operation $P \joino S$
eliminates the guarantee that $(cid,pid)$
functionally determines color in table $P$.
We similarly lose the fact that $(cid,sid)$ functionally determines $state$ in table $S$.

The anomaly is a result of a closed world assumption in relational databases.
We cannot finitely list that any $(cid,pid,sid)$ key in $P \joino S$ 
for which the subkey $(cid,pid)=(T,p01)$ has color $red$,
because there are an unlimited number of possible $sid$s
(recall that associative tables are \emph{total} functions, 
in the case of $sid$ from the domain of all possible strings).
We cannot use $red$ as the default value for color in $P \joino S$ either,
because other $(cid,pid)$ pairs in $P$ have color other than $red$, such as $(M,p01)$.

A closed-world solution requires that $S$ lists all the 
$(cid,sid)$ pairs that could be considered in queries.
This holds in many real world database scenarios, 
where we would not query for data not explicitly inserted into the database.

An alternative solution extends the definition of an associative table 
to allow \emph{default functions} instead of default values.
Evaluating a table $A$ at a certain key $k$ via $A(k)$
would lookup the key in the table, returning the looked-up value if present, 
and returning the result of a default function $f(k)$ if not present.
Database storage mechanisms would face greater challenge comparing tables for equality
and efficiently managing storage, especially when the default function is expensive.
On the other hand, we would gain greater expressiveness
for representing the structure inside an associative table.

Table normalization is a third solution: to not join tables $S$ and $P$ at all.
Instead we represent $S$ and $P$ separately in subsequent computations, 
which will faithfully maintain that $(cid,pid)$ determines color 
and $(cid,sid)$ determines state.
In fact there is plenty of good future work
in relating functional dependency and normal form concepts such as
inclusion, join, and multivalued dependencies.

\subsubsection{Automatic Promotion}
Having relaxed join to take the union of $A$ and $B$'s value attributes,
we now consider the case that some of $A$'s key attributes may be value attributes in $B$
and vice versa. 

A solution following the semantics of relational inner join is to 
``promote'' the value attributes of one table that are key attributes in the other 
to both have key attribute status.
In order to prevent losing support in the table whose value attribute becomes a key attribute
(call this table $P$),
we introduce a temporary ``indicator value attribute'' with 1s in rows from $P$'s support and 0s elsewhere.
Think: move the vertical bar separating key and value attribute to the right,
and leave a new column in its old place to the right of the bar with a marker indicating columns in support.

After the promotion, we follow the previous section's automatic multiply-by-one then strict join recipe,
and then remove the temporary indicator attribute.
See Figure~\ref{fAutoPromote} for an example.
The join effectively ``attaches'' the value attributes of $C$
to rows of $P$ where the `color' attribute matches, 
just as one would expect from relational inner join.

\begin{figure}
\centering
\subfloat[$P$]{ 
  \begin{tabular}{c|c}
   & [white]  \\
  pid & color \\
  \hline
  p01 & blue \\
  p02 & red  \\
  p03 & blue \\
  \end{tabular}
  \label{fAutoPromoteP}
} $\quad$
\subfloat[$\ext_{promote(color)}(P)$]{ 
  \begin{tabular}{cc|c}
   & & [0]  \\
  pid & color & ind \\
  \hline
  p01 & blue & 1 \\
  p02 & red  & 1 \\
  p03 & blue & 1 \\
  \end{tabular}
  \label{fAutoPromotePP}
} $\quad$
\subfloat[$C$]{ 
  \begin{tabular}{c|c}
   & [n] \\
  color & pretty\\
  \hline
  blue & y \\
  green & y \\
  \end{tabular}
  \label{fAutoPromoteC}
} $\quad$
\subfloat[$P \joino C =\newline \ext_{\text{supone}(pretty)}(\ext_{\text{promote}(color)}(P))\newline \sjoino C$]{ 
  \begin{tabular}{cc|c}
   & & [n]  \\
  pid & color & pretty \\
  \hline
  p01 & blue & y \\
  p03 & blue & y \\
  \end{tabular}
  \label{fAutoPromotePPC}
}
\caption{Automatic promotion in $P \joino C$; $f_1(pid,color)=color$}
\label{fAutoPromote}
\end{figure}


\section{Lara's Properties: Identities and Equivalences}

\subsection{Table Decomposition}
\label{sDecompose}
Any associative table $A$ can be decomposed into 
the union of single-value-attribute tables, as
\[ A = \Pi_{v_1}(A) \union \Pi_{v_2}(A) \union \dots \union \Pi_{v_n}(A) \]
where $v_1, \dots, v_n$ are $A$'s value attributes.

Decomposition allows programs to process multi-value-attribute associative tables 
using a composition of operations on single-value-attribute tables.
This is important for systems that cannot handle storing or processing 
more than one value attribute at a time.
Some column stores fall into this category.

One way to interpret the difference 
between the above left and right hand side is that
a table containing all $n$ value attributes is a row store, whereas
a set of $n$ tables with one value attribute each is a column store.
Hybrid stores have between 1 and $n$ tables, each with a number of value attributes
that sums to $n$.

Each store is suitable for different kinds of analytics.
Row storage (one table with $n$ value attributes) 
is more useful for maps which use all values as arguments.
Column storage ($n$ tables with 1 value attribute) 
is more useful for operations involving a single attribute at a time.

Both row and column storage options are expressible by an implementation 
as a result of the table decomposition equation above.

\subsection{Union and Join relative to Sum and Multiply}

Union and join inherit many properties of $\oplus$ and $\otimes$.
If $\oplus$ or $\otimes$ are associative, commutative, or idempotent, 
then so are $\uniono$ or $\joino$ respectfully.

If $\oplus$ has an identity $0$
and $A$'s value attributes all have 0 as their default value,
then the identity of $A$ under $\uniono$
is the table with all possible key attributes and no value attributes.

The identity of $\join$ is the table with no key attributes and no value attributes.
If $\otimes$ has an identity which we label $1$,
then the identity of $\sjoino$ is the table with no key attributes
and all possible value attributes, all with default value 1
and no other mappings.

The additive identity of $\uniono$ 
is necessarily the multiplicative annihilator of $\joino$.

The following sections illustrate two ways that join and union inherit 
the distributive property of $\otimes$ over $\oplus$.

\subsubsection{Conditions to Distribute Join over Union}

Suppose that $\otimes$ distributes over $\oplus$, such that 
$a \otimes (b \oplus c) = (a \otimes b) \oplus (a \otimes c)$.
Then join $\sjoino$ distributes over union $\uniono$, such that
$A \sjoino (B \uniono C) = (A \sjoino B) \uniono (A \sjoino C)$,
as long as $A$ and $B$ have no keys in common that are not also present in $C$,
and that $A$ and $C$ have no keys in common that are not also present in $B$.

\begin{figure}[hb]
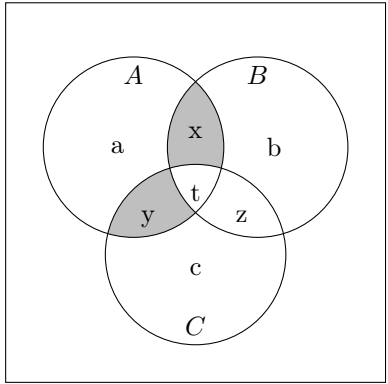

\centering
\begin{venndiagram3sets}[labelOnlyA={a},labelOnlyB={b},labelOnlyC={c},
labelOnlyAB={x},labelOnlyAC={y},labelOnlyBC={z},labelABC={t},
]
\fillACapBNotC
\fillACapCNotB 
\end{venndiagram3sets}
\caption{Visual depiction of keys of A, B, C}
\label{fVennABC}
\end{figure}

Proof: Suppose that associative tables 
$A$ has keys $a, x, y, t$;
$B$ has keys $b, x, z, t$; and
$C$ has keys $c, y, z, t$. 
This assortment of keys covers all combinations of sharing keys
between $A$, $B$, and $C$, as shown in Figure~\ref{fVennABC}.

We expand the definitions as follows.
Tick marks indicate variables under summation.
\begin{align*}
A(a,x,y,t) \sjoino \big( B(b,x,z,t) \uniono C(c,y,z,t) \big)
&= \big( A(a,x,y,t) \sjoino B(b,x,z,t) \big) \uniono \big( A(a,x,y,t) \sjoino C(c,y,z,t) \big) \\
A(a,x,y,t) \otimes \Big( \bigoplus_{b',x'} B(b',x',z,t) \oplus \bigoplus_{c',y'} C(c',y',z,t) \Big)
&= \bigoplus_{b'} \Big( A(a,x,y,t) \otimes B(b',x,z,t) \Big) \oplus \bigoplus_{c'} \Big( A(a,x,y,t) \otimes C(c',y,z,t) \Big) \\
A(a,x,y,t) \otimes \Big( \bigoplus_{b',x'} B(b',x',z,t) \oplus \bigoplus_{c',y'} C(c',y',z,t) \Big)
&= A(a,x,y,t) \otimes \Big( \bigoplus_{b'} B(b',x,z,t) \oplus \bigoplus_{c'}  C(c',y,z,t) \Big) 
\end{align*}
The left and right hand sides equal in the absence of keys $x$ and $y$.
The equality holds for values common to $A$ and $B$, 
values common to $A$ and $C$, and values common to $A$, $B$, and $C$
(other values do not exist in the answer, by definition of strict join).
$\qed$

\subsubsection{Pushing Union through Join}

The equivalence in this section is an application of the 
Generalized Distributive Law \cite{aji2000generalized} to Lara.
Suppose we have the query

\begin{centerverbatim}
select x.A, sum(z.D) from R x, S y, T z
where x.B = y.B and y.C = z.C 
group by x.A

\end{centerverbatim}

We model the query in Lara as the following. 
Each table has two key attributes named from the query and an indicator value attribute v,
whose values are either 1 to mark the presence of a key or 0.

\begin{figure}[htb]
\centering
\subfloat[R]{ 
  \begin{tabular}{cc|c}
   & & [0] \\
  A & B & v \\
  \hline
  $a_1$ & $b_1$ & 1 \\
  \vdots & \vdots & \vdots \\
  \end{tabular}
  \label{f1}
}
\subfloat[S]{ 
  \begin{tabular}{cc|c}
   & & [0] \\
  B & C & v \\
  \hline
  $b_1$ & $c_1$ & 1 \\
  \vdots & \vdots & \vdots \\
  \end{tabular}
  \label{f2}
}
\subfloat[T]{ 
  \begin{tabular}{cc|c}
   & & [0] \\
  C & D & v \\
  \hline
  $c_1$ & $d_1$ & $d_1$ \\
  \vdots & \vdots & \vdots \\
  \end{tabular}
  \label{f3}
}
\caption{Schema of example}
\label{f}
\end{figure}

The query's naive logical plan is 
\[ (R \join S \join T) \union_+ E_A \]

A smarter logical plan is 
\[ (((R \join S) \union_+ E_{A,C}) \join T) \union_+ E_A \]

We see the plan rewrite is correct by writing out
\begin{align*}
&((((R \join S) \union_+ E_{A,C}) \join T) \union_+ E_A)(a) \\
= &\sum_{c,d} \Big( \sum_b R(a,b) S(b,c) \Big) T(c,d) \\
= &\sum_{b,c,d} R(a,b) S(b,c) T(c,d) \quad \text{ (because * distributes over +)} \\
= &((R \join S \join T) \union_+ E_A)(a) 
\end{align*}

This is an instance of the more general equivalence
\[ (A \joino B) \uniono C 
= ((A \uniono E_{K_C \cup K_B}) \joino (B \uniono E_{K_C \cup K_A})) \uniono (C \uniono E_{K_A \cup K_B}) \]
assuming $\otimes$ distributes over $\oplus$.
The variables $K_A, K_B, K_C$ refer to the key attributes of $A, B, C$.
The expression $E_x$ denotes an ``empty table'' with key attributes $x$,
and value attributes and default values that agree with its usage's context.

\section{Translating Lara's and Other Algebras' Operations}
\label{sTranslateOperations}

\subsection{Relational Operations}
\label{sRelationalOperations}

Selection $\sigma_f(A)$ (where $f$ is a predicate)
is $\ext_{f'}(A)$,
where $f'(\bar{k},\bar{v}) = \bif f(\bar{k},\bar{v}) \bthen \bar{v} \belse \bar{0}$.
We set rows for which the predicate $f$ returns false to the default values
and leave other rows untouched.

Projecting away value attributes is a simple $\ext$ application in Lara.
Projecting away key attributes incurs aggregation and is done with $\uniono$.%
\footnote{Since the key attributes of an associative table roughly align with
the indexes or access path of a relational table, 
projecting away a key attribute of an associative table 
is like projecting away the primary key column of a relational table:
we don't usually physically project away the primary key because we use it
as part of the database access path. Instead the primary key remains in the table, 
just like the key attribute remains in the signature of the associative table when treated as a function.}

Cartesian product is a join between tables with no key attributes in common.

Aggregation (including `group by') and union are covered by table union.

Suppose $A$ has key attributes $k_1, k_2$; $B$ has key attributes $k_2$;
$A$ and $B$ have a value attribute called $v$;
and $B$'s contents of $v$ are 1s and 0s.
Relational difference $A \setminus B$, 
interpreted as removing the rows of $A$ for which $A$'s $k_2$ component matches 
a row in $B$'s support (more generally, all the keys in $K_A \cap K_B$ must match), 
is constructed via the following steps:
(1) $A \join B$ to select the rows that should be removed, 
(2) additive negation $\ext_{neg}(A \join B)$ 
where $neg(k_1, k_2, v) = -v$, and
(3) union $A \uniono \ext_{neg}(A \join B)$,
adding the original values of $A$
to the negative values of $A$ designated for removal.

Relational division $A \div B$ has a few different plausible definitions for associative tables.
One definition only involves the support of $A$ and $B$.  
We say that $A \div B = C$ if 
$K_C = K_A \setminus K_B$ and 
$\supp(C \join B) \subseteq \supp(A)$. 
For this definition, assuming that $A$ and $B$ have one value that is all 1s and 0s,
$A \div B = (A \join_* B) \union_+ E_{K_L \setminus K_R} == (B \union_+ E)$,
where == is shorthand for a $\ext$ that 
sets values equal to the scalar $B \union_+ E$ to 1 and the rest to 0.

The following subsections present a generalized form of relational division 
and the relational outer product.

\subsubsection{Generalized Relational Division}

\begin{figure}[htb]
\vspace{-2em}
\centering
\subfloat[$C$]{ 
  \begin{tabular}{l|l}
   & [0.0]  \\
  car & v \\
  \hline
  compact & 2.0 \\
  SUV & 5.0 \\
  electric & 1.0 \\
  \end{tabular}
}  $\;$
\subfloat[$P$]{ 
  \begin{tabular}{l|l}
   & [0.0]  \\
  fuel & v \\
  \hline
  reg & 2.0 \\
  prem & 3.0 \\
  \end{tabular}
} $\;$
\subfloat[$T := C \sjoin_* P$]{ 
  \begin{tabular}{ll|l}
   & & [0.0]  \\
  car & fuel & v \\
  \hline
  compact & reg  & 4.0 \\
  compact & prem & 6.0 \\
  SUV & reg & 10.0 \\
  SUV & prem & 15.0 \\
  electric & reg & 2.0 \\
  electric & prem & 3.0 \\
  \end{tabular}
} 

\subfloat[$p^{-1}_1$]{ 
  \begin{tabular}{l|l}
   & [0.0]  \\
  fuel & v \\
  \hline
  reg & 0.50 \\
  \end{tabular}
} $\;$
\subfloat[$(T \sjoin_* p_1^{-1}) \union E_{car}$]{ 
  \begin{tabular}{l|l}
   & [0.0]  \\
  car & v \\
  \hline
  compact & 2.0 \\
  SUV & 5.0 \\
  electric & 1.0 \\
  \end{tabular}
} $\;$
\subfloat[$p^{-1}_2$]{ 
  \begin{tabular}{l|l}
   & [0.0]  \\
  fuel & v \\
  \hline
  prem & 0.333 \\
  \end{tabular}
} $\;$
\subfloat[$(T \sjoin_* p_1^{-1}) \union E_{car}$]{ 
  \begin{tabular}{l|l}
   & [0.0]  \\
  car & v \\
  \hline
  compact & 2.0 \\
  SUV & 5.0 \\
  electric & 1.0 \\
  \end{tabular}
} $\;$
\subfloat[$T \div P$]{ 
  \begin{tabular}{l|l}
   & [0.0]  \\
  car & v \\
  \hline
  compact & 2.0 \\
  SUV & 5.0 \\
  electric & 1.0 \\
  \end{tabular}
} 
\caption{Division example, from Cartesian product}
\label{fDivCar}
\end{figure}

\begin{figure}[ht]
\vspace{-1em}
\centering
\subfloat[$P$]{ 
  \begin{tabular}{l|l}
   & [0.0]  \\
  fuel & v \\
  \hline
  reg & 2.0 \\
  prem & 3.0 \\
  \end{tabular}
} $\;$
\subfloat[$T$]{ 
  \begin{tabular}{ll|l}
   & & [0.0]  \\
  car & fuel & v \\
  \hline
  compact & reg  & 4.0 \\
  SUV & prem & 21.0 \\
  electric & reg & 3.0 \\
  electric & prem & 7.0 \\
  \end{tabular}
} 

\subfloat[$p^{-1}_1$]{ 
  \begin{tabular}{l|l}
   & [0.0]  \\
  fuel & v \\
  \hline
  reg & 0.50 \\
  \end{tabular}
} $\;$
\subfloat[$(T \sjoin_* p^{-1}_1) \union E_{car}$]{ 
  \begin{tabular}{l|l}
   & [0.0]  \\
  car & v \\
  \hline
  compact & 2.0 \\
  electric & 1.5 \\
  \end{tabular}
} $\;$
\subfloat[$p^{-1}_2$]{ 
  \begin{tabular}{l|l}
   & [0.0]  \\
  fuel & v \\
  \hline
  prem & 0.333 \\
  \end{tabular}
} $\;$
\subfloat[$(T \sjoin_* p^{-1}_1) \union E_{car}$]{ 
  \begin{tabular}{l|l}
   & [0.0]  \\
  car & v \\
  \hline
  SUV & 7.0 \\
  electric & 2.333 \\
  \end{tabular}
} $\;$
\subfloat[$T \div P$]{ 
  \begin{tabular}{l|l}
   & [0.0]  \\
  car & v \\
  \hline
  electric & 1.5 \\
  \end{tabular}
} 
\caption{Division example, no match to Cartesian product}
\label{fDivNoCar}
\end{figure}

\vspace{-1em}
Relational division is the inverse of relational multiplication, 
which is more commonly known as Cartesian product,
giving the identity $(A \times B) \div B = A$.
In the case of Lara, associative table division is the inverse of associative table join
between tables with disjoint keys.
We should define the semantics of associative table division 
to behave the same way on sets as relational division.
For cases beyond sets, associative table division should ``undo'' 
the multiplication inside the associative table join operation.

With the above criteria, we define associative table division 
as the ``biggest table $C$ whose keys are disjoint from $B$ such that $B \joino C \leq A$'':
\[A \div_\otimes B = \max\{ C \mid K_C \cap K_B = \emptyset \land C \joino B \leq A\}\]
The division is parameterized by a multiplication operation $\otimes$ which the division process inverts.
We require that the value attributes common to $A$ and $B$ have a partial order $\leq$
with the corresponding default value $0$ as the least element.
We also require that $\otimes$ is monotonically increasing, 
in the sense that $B \leq B' \Rightarrow A \joino B \leq A \joino B'$.

Computing the result of $A \div_\otimes B$ is easiest 
when the value attributes common to $A$ and $B$ have multiplicative inverses.
This occurs when the value type $\tau$ of $A$ and $B$'s share value attribute
forms a group with $\otimes$,
with multiplicative identity equal to the default value for both $A$ and $B$.
Like relational division, we also require that $K_B \subseteq K_A$, 
that the keys of $B$ are a subset of the keys of $A$.

We compute $A \div_\otimes B$ as follows, given that ($\tau, \otimes$) 
is a group with multiplicative inverse equal to the default value for each value attribute
and that $K_B \subseteq K_A$.
\begin{align*}
&A \div_\otimes B := \underset{b \in B}{\mathlarger{\mathlarger{\mathlarger{\sjoin}}}_{\min}} 
\big( (A \sjoino b\inv) \union E_{K_A \setminus K_B} \big) \\
&\text{ where } v\inv = \begin{cases}
0, \bif v = 0 \text{ (the default value)}\\
v\inv, \bif v \neq 0 \\
\end{cases} 
\end{align*}

The large join operator takes the strict join (with $\otimes = \min$)
of the expression $(A \sjoino b\inv) \union E_{K_A \setminus K_B}$
for every row $b$ in the support of $B$.
The $b\inv$ is shorthand for a map operation as defined above.
The notation $E_{K_A \setminus K_B}$
is the empty table with the keys of $A$ not in $B$.

Figures~\ref{fDivCar} and~\ref{fDivNoCar} show an example.
In Figure~\ref{fDivCar}, we verify an instance of the identity
$(C \times P) \div P = C$,
which in Lara corresponds to $(C \sjoino P) \div_\otimes P = C$.
The $\otimes$ is arithmetic multiplication over the non-negative real numbers.

An alternative way to formulate division that runs on all of $B$'s rows at once 
instead of using the ``big $\underset{b \in B}{\mathlarger{\mathlarger{\mathlarger{\sjoin}}}_{\min}}$'' operator
is to use an appended ``counter column'' to count the number of rows matched in $B$.
The following algorithm adds a counter column via $\map_a$, 
counts the number of matches while computing the join in $\union_{[\min,+]}$, 
and deletes rows in the result that did not match every row in $B$'s support with $\map_d$.
\begin{align*}
&A \div_\otimes B := \map_d( \map_a(A \sjoino B\inv) \union_{[\min,+]} E_{K_A \setminus K_B} ) \\
&\text{ where } a(k,v) = \begin{cases}
(0, 0), \bif v = 0 \\
(v, 1), \bif v \neq 0 \text{ (add a counter value)}\\
\end{cases} \\
&\text{ and } d(k,(v,i)) = \begin{cases}
0, \bif v < |B| \text{ (delete rows not matching all rows of $B$)} \\
v, \bif v = |B| \\
\end{cases} \\
&\text{ and } |B| = \map_f(B) \union_+ E \text{ with } f(k,v) = \begin{cases}
0, \bif v = 0 \\
1, \bif v \neq 0 \\
\end{cases} 
\end{align*}

We demonstrate the alternative algorithm on the example from Figure~\ref{fDivNoCar}
 in Figure~\ref{fDivNoCarAlt} which yields the same answer.
We use the name $i$ for the counter column.

\begin{figure}[ht]
\vspace{-1em}
\centering
\subfloat[$P$]{ 
  \begin{tabular}{l|l}
   & [0.0]  \\
  fuel & v \\
  \hline
  reg & 2.0 \\
  prem & 3.0 \\
  \end{tabular}
} $\;$
\subfloat[$T$]{ 
  \begin{tabular}{ll|l}
   & & [0.0]  \\
  car & fuel & v \\
  \hline
  compact & reg  & 4.0 \\
  SUV & prem & 21.0 \\
  electric & reg & 3.0 \\
  electric & prem & 7.0 \\
  \end{tabular}
} 

\subfloat[$X := \map_a(T \sjoin_* P^{-1})$]{ 
  \begin{tabular}{ll|ll}
   & & [0.0] & [0] \\
  car & fuel & v & i \\
  \hline
  compact & reg  & 2.0 & 1 \\
  SUV & prem & 7.0 & 1\\
  electric & reg & 1.5 & 1 \\
  electric & prem & 2.333 & 1 \\
  \end{tabular}
} 
\subfloat[$Y := X \union_{[\min,+]} E_{car}$]{ 
  \begin{tabular}{l|ll}
   & [0.0] & [0] \\
  car & v & i \\
  \hline
  compact & 2.0 & 1 \\
  SUV & 7.0 & 1 \\
  electric & 1.5 & 2 \\
  \end{tabular}
} 
\subfloat[$T \div P := \map_d(Y)$]{ 
  \begin{tabular}{l|l}
   & [0.0] \\
  car & v \\
  \hline
  electric & 1.5 \\
  \end{tabular}
}
\caption{Division example, no match to Cartesian product; alternate approach}
\label{fDivNoCarAlt}
\end{figure}

\subsubsection{Relational Outer Join}

Relational outer join is 
\[  A \fullouterjoin B = 
    (A \sjoin (\ext_{replace(V_A)}(B) \union E_{K_B \setminus K_A})) \union 
    (B \sjoin (\ext_{replace(V_B)}(A) \union E_{K_A \setminus K_B}))
\]
where $E_K$ is the table with key attributes $K$ and no value attributes,
and the function $replace(V)$ replace existing value attributes 
with the value attributes of $V$, 
with contents 1 everywhere in the original support and 0 elsewhere.
The missing $\oplus$ from the inner $\union$ operations indicates that
summed 1s remain as 1 as opposed to summing as integers.
No summing occurs in the outer $\union$ because the value attributes are disjoint.
See Figure~\ref{fOuterJoin} for an example.

The rows of $A \fullouterjoin B$ in Figure~\ref{fOuterJoinPZSZ} 
above the dashed line are the same rows resulting from inner join $A \join B$.
Rows below the dashed line are new rows introduced by outer join.
When $A$ and $B$ have no keys in common, 
outer join is equivalent to inner join
and we may rewrite $A \fullouterjoin B$ as $A \join B$.

\begin{figure}[tbh]
\centering
\subfloat[$P$]{ 
  \begin{tabular}{cc|c}
   & & [white]  \\
  cid & pid & color \\
  \hline
  M & p01 & blue \\
  T & p01 & red  \\
  M & p02 & green  \\
  W & p01 & yellow \\
  \end{tabular}
  \label{fOuterJoinP}
}
\subfloat[$P \sjoin (\ext_{replace(V_P)}(S) \union E_{K_S \setminus K_P})$]{ 
  \begin{tabular}{ccc|c}
   & & & [white] \\
  cid & pid & sid & color\\
  \hline
  M & p01 & s01 & blue  \\
  M & p01 & s02 & blue  \\
  M & p02 & s01 & green \\
  M & p02 & s02 & green \\
  T & p01 & s01 & red   \\
  T & p01 & s02 & red   \\
  W & p01 & s01 & yellow\\
  W & p01 & s02 & yellow\\
  \end{tabular}
  \label{fOuterJoinPZ}
}
\subfloat[$S$]{ 
  \begin{tabular}{cc|c}
   & & [GA] \\
  cid & sid & state \\
  \hline  
  M & s01 & WA \\
  M & s02 & NJ \\
  T & s02 & DE \\
  F & s01 & CA \\
  \end{tabular}
  \label{fOuterJoinS}
}
\subfloat[$S \sjoin (\ext_{replace(V_S)}(P) \union E_{K_P \setminus K_S})$]{ 
  \begin{tabular}{ccc|c}
   & & & [GA] \\
  cid & pid & sid & state\\
  \hline
  M & p01 & s01 & WA  \\
  M & p01 & s02 & NJ  \\
  M & p02 & s01 & WA \\
  M & p02 & s02 & NJ \\
  T & p01 & s02 & DE   \\
  T & p02 & s02 & DE   \\
  F & p01 & s01 & CA\\
  F & p02 & s01 & CA\\
  \end{tabular}
  \label{fOuterJoinSZ}
}

\subfloat[$(P \sjoin (\ext_{replace(V_P)}(S) \union E_{K_S \setminus K_P})) \union (S \sjoin (\ext_{replace(V_S)}(P) \union E_{K_P \setminus K_S}))$]{ 
  \begin{tabular}{ccc|cc}
   & & & [white] & [GA] \\
  cid & pid & sid & color & state \\
  \hline
  M & p01 & s01 & blue & WA \\
  M & p01 & s02 & blue & NJ \\
  M & p02 & s01 & green & WA \\
  M & p02 & s02 & green & NJ \\
  T & p01 & s02 & red & DE  \\
  \hdashline
  T & p01 & s01 & red & (GA)  \\
  T & p02 & s02 & (white) & DE  \\
  W & p01 & s01 & yellow & (GA)  \\
  W & p01 & s02 & yellow & (GA)  \\
  F & p01 & s01 & (white) & CA  \\
  F & p02 & s01 & (white) & CA  \\
  \end{tabular}
  \label{fOuterJoinPZSZ}
}
\caption{Example of outer join of $P$ with $S$}
\label{fOuterJoin}
\end{figure}

\subsection{CombBLAS Array Operations}


\begin{figure}[ht]
\centering
\includegraphics[width=13cm]{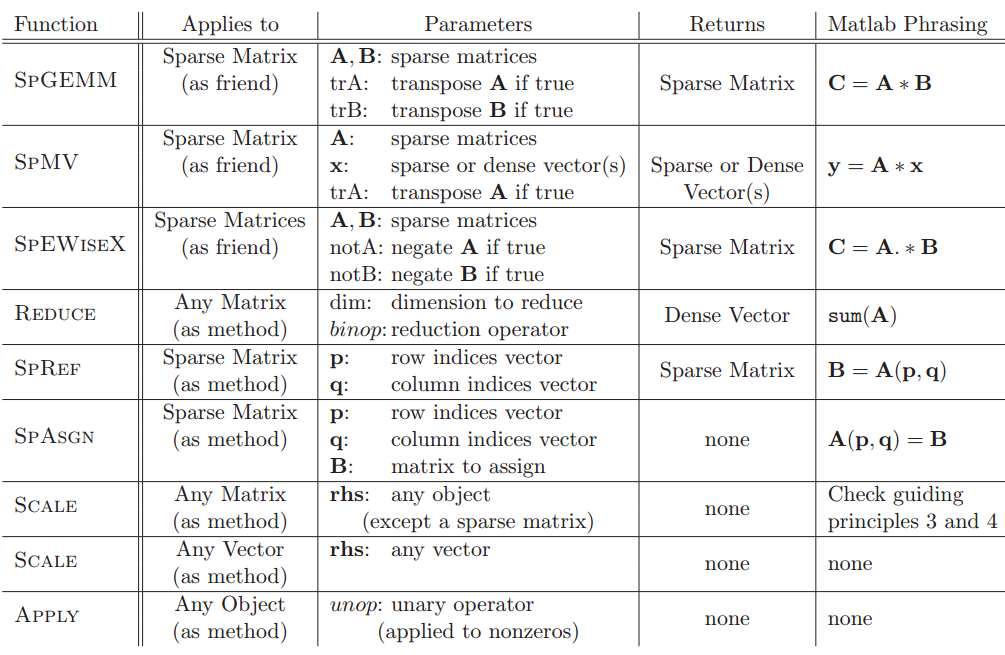}
\caption{CombBLAS API \cite{bulucc2011combinatorial}}
\label{fCombBLAS}
\end{figure}

The Combinatorial BLAS (CombBLAS) is a C++ library for distributed sparse matrix computation \cite{bulucc2011combinatorial}.
Its API, shown in Figure~\ref{fCombBLAS}, is representative of many operations we perform in array or linear algebra.
In fact, CombBLAS is one of the foundation implementations guiding the design of the GraphBLAS specification, 
a standard for primitive graph algorithm building blocks in the language of linear algebra \cite{mattson2014standards}.
Other BLAS-like implementations include ScaLAPACK \cite{blackford1997scalapack}, SciDB \cite{stonebraker2013scidb}, and Graphulo \cite{hutchison2015graphulo}.

CombBLAS restricts its attention to one-dimensional vectors and two-dimensional matrices.
Because Lara can model any-dimensional tensors, most translations from CombBLAS to Lara operations 
are special cases of Lara operations.

The CombBLAS SpGEMM operation stands for sparse generalized matrix-matrix multiplication.
It has two flags, trA and trB, which indicate whether $A$ or $B$ should be transposed
before the multiplication.
We need not model the trA and trB flags in Lara explicitly because 
they can be written as Lara rename operations (that is, 
a sequence of ext operations that swap the names of the two key attributes of $A$ or $B$).

CombBLAS SpMV stands for sparse matrix-vector multiplication, with separate versions for the sparse or dense case of $v$.
Because Lara is a logical algebra, we can defer sparse/dense specialization concerns to physical implementations.

A Lara pattern for SpGEMM is $(A \sjoino B) \uniono E_{r,c}$.
Tables $A$ and $B$ must have exactly two key attributes each,
one of which must be the same.
For example $A$ could have key attributes $r$ and $m$
and $B$ could have key attributes $m$ and $c$.
The resulting table has key attributes $r$ and $c$.

Similarly, a Lara pattern for SpMV is $(A \sjoino v) \uniono E_r$.
The table $v$ representing a vector has only one key attribute,
which must match of $A$'s two key attributes.
If $A$ has keys $r$ and $c$, for example, then $(A \sjoino v) \uniono E_r$ has key $r$.

CombBLAS SpEWiseX stands for sparse element-wise multiplication,
which multiplies values at matching positions.
A Lara pattern for SpEWiseX is $A \sjoino B$
when tables $A$ and $B$ have the same key attributes.

The CombBLAS SpEWiseX signature has two negation flags `notA' and `notB'.
If notB is set, then the SpEWiseX operation results in $A$ with entries that $B$ has set zeroed out.
Setting both notA and notB does not make sense and is disallowed.

The notB option with SpEWiseX has the same effect as relational difference $A \setminus B$.
We can implement this in Lara following the $A \uniono \ext_{neg}(A \join B)$ recipe
from Section~\ref{sRelationalOperations}.


CombBLAS Reduce reduces the dimension of a matrix by summing along rows or columns with a given $\oplus$ operator.
This is exactly equivalent to Lara $A \uniono E_X$,
where $X$ is the set of key attributes we wish to retain (not reduce) in the output.

CombBLAS SpRef extracts a subset of a matrix based on aligned index vectors $p$ and $q$.
We can model SpRef in Lara by two methods: $\map_f(A)$ and $A \join R$.
Function $f$ in the first method embeds index vectors $p$ and $q$ 
such that $f$ zeros out values in all entries except the entries we wish to retain.
Table $R$ in the second method has the same key attributes of $A$,
has support equal to the entries we wish to retain from $p$ and $q$,
and has values all equal to 1.
The $\otimes$ in the $\join$ is any multiplication operator that 
respects 1 as its identity and 0 as its annihilator. 

CombBLAS SpAsgn assigns the values of a matrix $B$ to a submatrix of $A$ based on aligned index vectors $p$ and $q$.
It is easiest to imagine SpAsgn in Lara as $\map_f(A)$, where
function $f$ updates entries in $A$ at locations determined by $p$ and $q$ with new values from $B$.
In the special case that the indices of $B$ match the indices we wish to update in $A$,
then we can write an update as $(A \union_{\oplus_{\setminus}} B) \union_+ B$,
which zeros out the elements of $A$ we want to update, then adds in the values of $B$
to replace the zeros where we want to update.

CombBLAS scale multiplies every row or column of a matrix $A$ by the matching entry in a vector $v$.
This is equivalent to the matrix multiplication $AD$ when scaling columns or $DA$ when scaling rows,
where $D$ is a diagonal matrix with vector $v$ on the main diagonal (\texttt{diag(v)}, in Matlab notation).

A more efficient way to implement Scale in Lara than the diagonal matrix multiplication is $A \sjoino v$.
Vector $v$ is a table with one key attribute that matches one of $A$'s key attributes.
These semantics multiply every ``column'' (or row) of $A$ by the element in the corresponding
``row vector'' (or column vector) $v$, depending on the schemas (names of key attributes) of $A$ and $v$.
Zero entries of $v$ multiply corresponding subsets of $A$ by zero, eliminating them.

CombBLAS treats scaling by a sparse vector with zero entries in a special way:
a zero entry in the vector indicates the corresponding subset of $A$ should not be scaled 
(as if entries in the subset of $A$ were multiplied by 1, but without the overhead of materializing 1s in the vector $v$).
We can mirror the semantics of CombBLAS Scale on a sparse vector $v$
by setting the default value of $v$ as an associative table to 1 instead of 0.
Appendix~\ref{aJoinZeros} explains how the Lara join operation is well-defined in this case 
despite the default value not acting as an annihilator. 

CombBLAS Apply applies a function to each element of a matrix.
This is Lara $\map$.

An additional operation not listed in \ref{fCombBLAS} but common to matrix operations
is SpEWiseSum: the element-wise sum of two sparse matrices, written as $A + B$. SpEWiseSum differs from SpEWiseX
in that the support of the resulting matrix (that is, the entries that are nonzero in the resulting matrix)
is the union of the supports of $A$ and $B$ rather than the intersection (ignoring cases when two nonzero entries sum to zero).
In Lara, SpEWiseSum is a $\union_+$ operation.

\subsection{Discrete Convolution}

A matrix convolution computes a new value to each entry 
based on a function from the values at ``nearby'' entries
determined by a pattern called a kernel.
For example, a vector blur convolution replaces each entry at position $i$ 
with the average of adjacent entries at position $i-1$ and $i+1$.
A ``prefix sum'' convolution replaces each entry with the sum of that entry's value
and values at entries preceding it.
Convolutions of one, two, and higher dimensions are an important family of computation
in image processing, machine learning, and many more disciplines.

We now formalize a description of a convolution's kernel for associative tables.
Let $\oplus$ be the convolution function.
Let $k$ be a value of an associative table's keys.
Let $D_k$ be the set of keys for which the result of the convolution at $k$ depends on.
Let $D\inv_k$ be the set of keys whose value resulting from the convolution depends on $k$.
For example, the convolution mapping each entry's value to the sum of the entry's
and preceding entry's value has $D_k = \{k-1, k\}$, $D\inv_k = \{k+1, k\}$, 
and $\oplus = +$.

In order for convolution to be well-defined on an associative table $A$,
we require that $|D\inv_k|$, the number of entries affected by $A(k)$, is finite. If $|D\inv_k|$ is infinite,
then the result of the convolution may not have finite support.

Depending on the definition of $D\inv_k$,
we have two methods to compute convolution in Lara.
The first requires that $D\inv_k$ is independent of $\supp(A)$;
the second requires that $D\inv_k \subseteq \supp(A)$.
If you know of a useful convolution whose $D\inv_k$ escapes both requirements 
yet has finite support, please contact the author.

Both methods do \emph{not} represent $D\inv_k$ in ordinal terms
as in $D\inv_k = \inf \{ k' > k | A(k') \neq 0 \}$,
which is the smallest key greater than $k$ in the support of $A$, if one exists.
Instead we represent $D\inv_k$ in absolute terms, as in $D\inv_k = \{ k+1 \} \cap \supp(A)$,
which is the key $k+1$ if it is in the support of $A$, and otherwise no key.

\subsubsection{Convolution when $D^{-1}_k$ does not depend on $A$'s support}

\begin{figure}[htb]
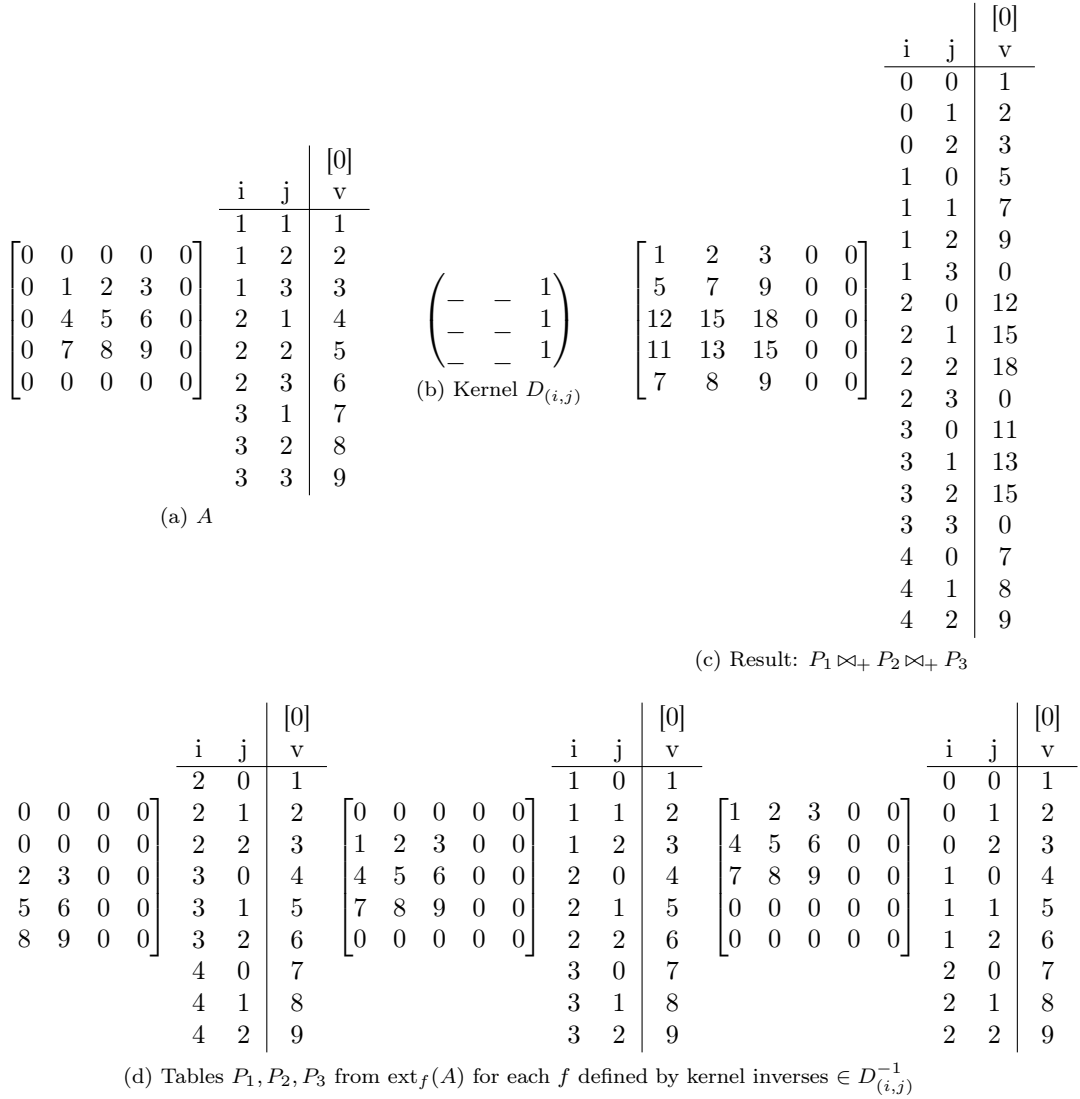

\centering
\subfloat[$A$]{ 
  $\begin{bmatrix}
  0 & 0 & 0 & 0 & 0 \\
  0 & 1 & 2 & 3 & 0 \\
  0 & 4 & 5 & 6 & 0 \\
  0 & 7 & 8 & 9 & 0 \\
  0 & 0 & 0 & 0 & 0 \\
  \end{bmatrix}$
  \begin{tabular}{cc|c}
   & & [0] \\
  i & j & v \\
  \hline
  1 & 1 & 1 \\
  1 & 2 & 2 \\
  1 & 3 & 3 \\
  2 & 1 & 4 \\
  2 & 2 & 5 \\
  2 & 3 & 6 \\
  3 & 1 & 7 \\
  3 & 2 & 8 \\
  3 & 3 & 9 \\
  \end{tabular}
  \label{fConvA}
}
$\quad$
\subfloat[Kernel $D_{(i,j)}$]{ 
  $\begin{pmatrix}
  \_ & \_ & 1 \\
  \_ & \_ & 1 \\
  \_ & \_ & 1
  \end{pmatrix}$
  \label{fConvK}
}
$\quad$
\subfloat[Result: $P_1 \join_+ P_2 \join_+ P_3$]{ 
  $\begin{bmatrix}
  1 & 2 & 3 & 0 & 0 \\
  5 & 7 & 9 & 0 & 0 \\
  12& 15& 18& 0 & 0 \\
  11& 13& 15& 0 & 0 \\
  7 & 8 & 9 & 0 & 0 \\
  \end{bmatrix}$
  \begin{tabular}{cc|c}
   & & [0] \\
  i & j & v \\
  \hline
  0 & 0 & 1 \\
  0 & 1 & 2 \\
  0 & 2 & 3 \\
  1 & 0 & 5 \\
  1 & 1 & 7 \\
  1 & 2 & 9 \\
  1 & 3 & 0 \\
  2 & 0 & 12\\
  2 & 1 & 15\\
  2 & 2 & 18\\
  2 & 3 & 0 \\
  3 & 0 & 11\\
  3 & 1 & 13\\
  3 & 2 & 15\\
  3 & 3 & 0 \\
  4 & 0 & 7 \\
  4 & 1 & 8 \\
  4 & 2 & 9 \\
  \end{tabular}
  \label{fConvR}
}

\subfloat[Tables $P_1, P_2, P_3$ from $\ext_f(A)$ for each $f$ defined by kernel inverses $\in D^{-1}_{(i,j)}$]{ 
  $\begin{bmatrix}
  0 & 0 & 0 & 0 & 0 \\
  0 & 0 & 0 & 0 & 0 \\
  1 & 2 & 3 & 0 & 0 \\
  4 & 5 & 6 & 0 & 0 \\
  7 & 8 & 9 & 0 & 0 \\
  \end{bmatrix}$
  \begin{tabular}{cc|c}
   & & [0] \\
  i & j & v \\
  \hline
  2 & 0 & 1 \\
  2 & 1 & 2 \\
  2 & 2 & 3 \\
  3 & 0 & 4 \\
  3 & 1 & 5 \\
  3 & 2 & 6 \\
  4 & 0 & 7 \\
  4 & 1 & 8 \\
  4 & 2 & 9 \\
  \end{tabular}
  $\begin{bmatrix}
  0 & 0 & 0 & 0 & 0 \\
  1 & 2 & 3 & 0 & 0 \\
  4 & 5 & 6 & 0 & 0 \\
  7 & 8 & 9 & 0 & 0 \\
  0 & 0 & 0 & 0 & 0 \\
  \end{bmatrix}$
    \begin{tabular}{cc|c}
   & & [0] \\
  i & j & v \\
  \hline
  1 & 0 & 1 \\
  1 & 1 & 2 \\
  1 & 2 & 3 \\
  2 & 0 & 4 \\
  2 & 1 & 5 \\
  2 & 2 & 6 \\
  3 & 0 & 7 \\
  3 & 1 & 8 \\
  3 & 2 & 9 \\
  \end{tabular}
  $\begin{bmatrix}
  1 & 2 & 3 & 0 & 0 \\
  4 & 5 & 6 & 0 & 0 \\
  7 & 8 & 9 & 0 & 0 \\
  0 & 0 & 0 & 0 & 0 \\
  0 & 0 & 0 & 0 & 0 \\
  \end{bmatrix}$
    \begin{tabular}{cc|c}
   & & [0] \\
  i & j & v \\
  \hline
  0 & 0 & 1 \\
  0 & 1 & 2 \\
  0 & 2 & 3 \\
  1 & 0 & 4 \\
  1 & 1 & 5 \\
  1 & 2 & 6 \\
  2 & 0 & 7 \\
  2 & 1 & 8 \\
  2 & 2 & 9 \\
  \end{tabular}
  \label{fConvP}
}

\caption{Matrix Convolution Example}
\label{fConv}
\end{figure}

Suppose we have a matrix $A$ and its associative table representation in Figure~\ref{fConvA},
and suppose we want to convolute $A$ by the kernel depicted in matrix form in Figure~\ref{fConvK}
with arithmetic sum as the kernel function.
The operation replaces each entry of $A$ at key $(i,j)$ with the sum of the entries at keys
$D_{(i,j)} = (i-1,j+1), (i,j+1), (i+1,j+1)$.
Notice that $D\inv_{(i,j)} = (i+1,j-1), (i,j-1), (i-1,j-1)$ is independent of $\supp(A)$.

We can implement the convolution operation in Lara by 
(1) for $k' \in D\inv_k$, use $\ext$ to construct tables with keys shifted according to $k'$
(illustrated with matrices and tables in Figure~\ref{fConvP}), and 
(2) joining together the resulting tables via $\joino$, using the kernel function as $\otimes$ (Figure~\ref{fConvR}).
The join operations are well-defined for $\otimes$ operations that behave more like $\oplus$
(in the sense that $a \oplus 0$ need not necessarily equal 0)
because we meet the $K_A = K_B$ condition defined in Section~\ref{aJoinZeros}.

We can implement any convolution operation whose kernel is independent of $A$'s support following the above method.
Even a prefix sum 1-D convolution method works, although $|D\inv_k| = |\supp(A)|$.

Our method for convolving tables takes inspiration from 
the implementation of convolution as BLAS matrix multiplication \cite{chellapilla2006high}.

\subsubsection{Convolution when $D^{-1}_k$ depends on and is a subset of $A$'s support}

A common operation on time-series data is to find the $d$-moving-sum.
This operation replaces each \emph{nonzero} value at time $t$ from an associative table $A$
with the sum of values in the time range $[t-d,t]$.
This is a convolution with kernel $D_t = [t-d,t]$.

We emphasize that the $d$-moving-sum must act on nonzero values
(that is, entries in the support of $A$)
because otherwise $D\inv_t = [t,t+d]$ has infinite support.
The restriction sets $D\inv_t = [t, t+d] \cap \supp(A)$ which is finite.
Notice that $D\inv_t \subseteq \supp(A)$.

For a concrete example, consider the table of time series data $T$ in Figure~\ref{fkwinT}
consisting of times $t$ and values $v$, and let $d=2$.
Then the series of operations in Figure~\ref{fkwin}
lead to the result in Figure~\ref{fkwinRTA}.

\begin{figure}[htb]
\centering
\subfloat[$T$]{ 
  \begin{tabular}{c|c}
   & [0]  \\
  t & v \\
  \hline
  1.0 & 4 \\
  1.3 & 8 \\
  2.5 & 6 \\
  3.1 & 2 \\
  5.0 & 3 \\ 
  9.0 & 42 \\
  \end{tabular}
  \label{fkwinT}
} $\;$
\subfloat[$T_0 := \newline \map_{f}(T)$]{ 
  \begin{tabular}{c|c}
   & [0.0]  \\
  t & v \\
  \hline
  1.0 & 1.0 \\
  1.3 & 1.3 \\
  2.5 & 2.5 \\
  3.1 & 3.1 \\
  5.0 & 5.0 \\ 
  9.0 & 9.0  \\
  \end{tabular}
  \label{fkwinT0}
} $\;$
\subfloat[$T'_0 := \newline \rho_{t\to t'}T_0$]{ 
  \begin{tabular}{c|c}
   & [0.0]  \\
  t' & v \\
  \hline
  1.0 & 1.0 \\
  1.3 & 1.3 \\
  2.5 & 2.5 \\
  3.1 & 3.1 \\
  5.0 & 5.0 \\ 
  9.0 & 9.0  \\
  \end{tabular}
  \label{fkwinT0'}
} $\;$
\subfloat[$R := T_0 \sjoin_{\otimes} T'_0$]{ 
  \begin{tabular}{cc|c}
   & & [0]  \\
  t &  t' & v \\
  \hline
  1.0 & 1.0 & 1 \\
  1.0 & 1.3 & 1 \\
  1.0 & 2.5 & 1 \\
  1.3 & 1.3 & 1 \\
  1.3 & 2.5 & 1 \\
  1.3 & 3.1 & 1 \\
  2.5 & 2.5 & 1 \\
  2.5 & 3.1 & 1 \\
  3.1 & 3.1 & 1 \\
  3.1 & 5.0 & 1 \\
  5.0 & 5.0 & 1 \\ 
  9.0 & 9.0 & 1  \\
  \end{tabular}
  \label{fkwinR}
} $\;$
\subfloat[$R \sjoin T$]{ 
  \begin{tabular}{cc|c}
   & & [0]  \\
  t &  t' & v \\
  \hline
  1.0 & 1.0 & 4 \\
  1.0 & 1.3 & 4 \\
  1.0 & 2.5 & 4 \\
  1.3 & 1.3 & 8 \\
  1.3 & 2.5 & 8 \\
  1.3 & 3.1 & 8 \\
  2.5 & 2.5 & 6 \\
  2.5 & 3.1 & 6 \\
  3.1 & 3.1 & 2 \\
  3.1 & 5.0 & 2 \\
  5.0 & 5.0 & 3 \\ 
  9.0 & 9.0 & 42 \\
  \end{tabular}
  \label{fkwinRT}
} $\;$
\subfloat[$(R \sjoin T) \union_+ E_{t'}$]{ 
  \begin{tabular}{c|l}
   & [0]  \\
  t' & v \\
  \hline
  1.0 & 4 \\
  1.3 & 4+8 \\
  2.5 & 4+8+6 \\
  3.1 & 8+6+2 \\
  5.0 & 2+3 \\
  9.0 & 42 \\
  \end{tabular}
  \label{fkwinRTA}
}
\caption{2-Moving-Sum Convolution Example}
\label{fkwin}
\end{figure}

The operations used in Figure~\ref{fkwin} are 
\begin{align*}
f(t,v) &:= \bif v=0 \bthen 0.0 \belse t \\
v \otimes v' &:= \bif v \neq 0 \land t \leq t' \leq t+2.0 \bthen 1 \belse 0
\end{align*}

\section{Related Work}

Insight for the current presentation of the join and union operations 
follows from three bodies of work:
\begin{enumerate}
\item Relational lattice operations \cite{spight2006first},
  which reduces the five standard relational operators
  to generalized union ($\union$) and natural join ($\bowtie$).
\item Multi-set relational algebra \cite{grefen1994multi},
  which treats tuples as arguments of a function
  that returns a natural number, 
  representing the number of occurrences of that tuple.
  We treat tuples as arguments of a function that
  reuturns an element of $V$.
\item The GraphBLAS API \cite{mattson2014standards} and the algebra of associative arrays \cite{kepner2015unified}, 
  both of which track the structures of abstract algebra.
\end{enumerate}


\section{Further Work}
\label{sFurtherWork}
This document describes Lara's core data structure and operations.
There are several additions Lara ought to have in order to be a viable language.
\begin{enumerate}
\item A comprehension syntax front-end.
\item Datalog-like variable reference syntax.
For the Part-Supplier-Request database of Figure~\ref{fPartSupplierRequest},
we might write $P(pid; \_, totWgt) \sjoin_* R(sid, pid; totWgt, \_)$
which concisely performs renaming, projection, and join 
to calculate the total weight of parts requested, grouped by requesting suppliers.
The syntax $A(\vec{k}; \vec{v})$ refers to keys $\vec{k}$ and values $\vec{v}$ from $A$.
We might write an expression in place of an attribute name to indicate an $\ext$ operation creating that attribute.
\item Variable assignment, in order to remember and reuse intermediary results.
\item Iteration or recursion.
\end{enumerate}
Adding these features would pave a path for programmers to write Lara expressions directly,
whether as a standalone query language or as a DSL embedded into existing general-purpose programming languages.
We anticipate programmers would find writing an algorithm in Lara easier
when an algorithm spans multiple families of computational systems,
which normally require stitching together pieces in different query languages.

A major test for the Lara algebra's usefulness
is to measure how well Lara performs in her role to facilitate translation 
between relational, array, graph, and key-value algebras.
Polystore optimization is quickly maturing;
we look forward to seeing how well she connects these algebras
and all the theorems and algorithms behind them.

\bibliographystyle{IEEEtran}

\bibliography{10_bibliography}

\clearpage
\appendix
\section{Lara Cheat Sheet}
\label{aLang}

A record is a tuple with a name for each component. 
The type of a record $\bar{r}$ is 
$\tau_{\bar{r}} = (h_1: \tau_1, \dots, h_n: \tau_n)$
where $h_1, \dots, h_n$ are unique string names we call headers that identify fields
and $\tau_1, \dots, \tau_n$ are the types of each field.

For example, we write a value of record type $(\text{age}: \mathbb{N}, \text{ score}: \mathbb{R})$
as $(\text{age}: 35, \text{ score}: 98.2)$, or when the header names are clear from context
we omit them as in $(35, 98.2)$.

Let $\pi_{H}$ denote the projection of a record to a subset of its components,
namely, the components whose name is in the set $H$.
We relax notation such that when we write $H$ as a single name, 
we interpret it as a singleton set.
We use the period symbol `.' for concatenation of records \emph{with disjoint headers}.

An associative table $A$
is a total function from a record of keys $\bar{K}$ to a record of values $\bar{V}$
with finite support and the requirement that $\bar{K}$ and $\bar{V}$ have disjoint headers.
The form of $A$ is a table listing mappings from $\bar{k}$ to $\bar{v}$.
(Small note: upper-case $\bar{K}$ is a record type; lower-case $\bar{k}$ is a record value.)
Keys $\bar{k}$ that do not appear in the table map to a default value $\bar{0}$.
We write the type of an associative table $A$ as
\[ A : [[ \bar{K} \mapsto \bar{V} : \bar{0} ]] \]
We call $A$ as a function via the expression $A(\bar{k})$, 
usually dropping nested parentheses for readability.

Suppose we have the following types and values (attribute types omitted) 
\begin{table}[h]
\centering
\begin{tabular}{lll}
$\bar{K}_A = (k_a, k_c)$ & $\bar{V}_A = (v_x, v_z)$ & $\bar{0}_A = (v_x: 0_x, v_z: 0_z)$ \\
$\bar{K}_B = (k_c, k_b)$ & $\bar{V}_B = (v_z, v_y)$ & $\bar{0}_B = (v_z: 0_z, v_y: 0_y)$ \\
\multicolumn{3}{c}{$A : [[\bar{K}_A \mapsto \bar{V}_B:\bar{0}_A ]]$} \\
\multicolumn{3}{c}{$B : [[\bar{K}_B \mapsto \bar{V}_B:\bar{0}_B ]]$} \\
\end{tabular}
\end{table}

\noindent
The following operations have type and result:
\begin{align*}
&A \union_{(\oplus_x,\oplus_z,\oplus_y)} B : [[k_c \mapsto (v_x, v_z, v_y): (0_x, 0_z, 0_y)]] \\
&(A \union_{(\oplus_x,\oplus_z,\oplus_y)} B)(c) :=
  \Big( 
    v_x: \bigoplus_{a}\!{}_x \: \pi_x A(a,c),\; 
    v_z: \bigoplus_{a}\!{}_z \: \pi_z A(a,c) \oplus_z \bigoplus_{b}\!{}_z \: \pi_z B(c,b),\;
    v_y: \bigoplus_{b}\!{}_y \: \pi_y B(c,b)
  \Big) \\
&\quad \text{requiring } \forall i, \oplus_i; 0_i \oplus_i i = i \oplus_i 0_i = i 
\qquad\qquad \text{($A$ and $B$'s $0_z$ must agree)} \\
\\
&A \sjoin_{\otimes_z} B : [[(k_a, k_c, k_b) \mapsto v_z : (0_z \otimes_z 0_z)]] \\
&(A \sjoin_{\otimes_z} B)(a,c,b) := (v_z: \pi_z A(a,c) \otimes_z \pi_z B(c,b)) \\
&\quad \text{requiring } \forall z; 0_z \otimes_z z = z \otimes_z 0_z = 0_z \otimes_z 0_z 
\quad \text{($A$ and $B$'s $0_z$ may differ in general)} \\
\\
&\ext_f(A) : [[ \bar{K}_A . \bar{K}' \mapsto \bar{V}' : \bar{0}' ]] \\
&\ext_f(A)(\bar{k}_A . \bar{k}') := f(\bar{k}_A . A(\bar{k}_A))(\bar{k}') \\
&\quad \text{where } f : \bar{K}_A \times \bar{V}_A \to (\bar{K}' \to \bar{V}') \\ 
&\quad \text{requiring } \forall \bar{k}_A, \bar{k}'; f(\bar{k}_A, \bar{0}_A)(\bar{k}') = \bar{0'} \text{ (constant) and }
\forall \bar{k}_A, \bar{v}_A; f(\bar{k}_A . \bar{v}_A) \text{ has finite support} \\
\\
&\text{\underline{Derived} --- Relaxed Join and Promote} \\
&A \join_{\otimes_z} B : [(k_a, k_c, k_b) \mapsto (v_x, v_z, v_y) : (0_x, 0_z \otimes_z 0_z, 0_y)]] \\
&A \join_{\otimes_z} B := \map_{\intro_{A,v_y}}(A) \sjoin_{\otimes_z} \map_{\intro_{B,v_x}}(B) \\
&\quad \text{where } \intro_{A,H} : \bar{K}_A \times \bar{V}_A \to \bar{V}_A . (H: \mathbb{B}) \quad\text{ ($H$ is disjoint from header of $K_A$ and $V_A$)} \\
&\quad \text{\phantom{where} } \intro_{A,H}(\_, \bar{v}_A) := \bif \bar{v}_A = \bar{0}_A \bthen \bar{0}_A . (H: 0) \belse \bar{v}_A . (H: 1) \\
&\quad \text{(auto-promotion: relaxed join first promotes values in $A$ with the same name as a key in $B$ and vice versa)} 
\\
&\promote_{v_x}(A) : [[ (k_a, k_c, v_x) \mapsto (v'_x, v_z) : (0, 0_z) ]] \\
&\promote_{v_x}(A)(a,c,x) := \ext_f(A) \text{ where } f(a,c,x,z) := \lambda(x'). \bif x = x' \bthen (v'_x: 1, v_z: z) \belse (v'_x: 0,v_z: 0_z) \\
&\quad \text{and the new name $v'_x$ is the old name $v_x$ concatenated with enough apostrophes to guarantee freshness} \\
\end{align*}

\clearpage
\section{Behavior of Join when Zero does not fully Annihilate}
\label{aJoinZeros}

This is motivated by the sparse vector case of the Scale operation in the CombBLAS API.
This version of the CombBLAS Scale function takes a sparse matrix $A$
and a sparse row vector $B$ and, for each \emph{nonzero} element of $B$ (call it $b$),
it multiplies each element in the corresponding column of $A$ by $b$.
Columns of $A$ corresponding to a zero element of $B$ are unmodified, as if they were multiplied by 1.

The join $A \sjoino v$ could capture the behavior of CombBLAS Scale in the sparse vector case
if we make the default value of $B$ the multiplicative identity 1.
However, joining $A$ (default value 0) with $B$ (default value 1)
would violate the constraints on $\otimes$ that,
for default value $0_A$ and $0_B$ from $A$ and $B$,
$\forall a; a \otimes 0_B = 0_A \otimes 0_B$ and
$\forall b; 0_A \otimes b = 0_A \otimes 0_B$.
The specific violation is that $a \otimes 1 = a \neq 0 \otimes 1 = 0$.

The constraints on $\otimes$ exist in order to ensure that $A \sjoino B$
has finite support, and they are necessary in general.
The constraints are not necessary for certain cases of $A$ and $B$.
The Scale operation above illustrates one such case: 
when the keys of $B$ are a subset of th keys of $A$: $K_B \subset K_A$.
This case does maintain finite support of $A \sjoino B$
since the condition $\forall a; a \otimes 0_B = 0_A \otimes 0_B$ is satisfied.

Table~\ref{tZeroConstraints} illustrates all the special cases of $A \sjoino B$ concerning keys.
The rows are different cases of relationships between the keys of $A$ and $B$:
equal keys, $A$'s keys a strict subset of $B$'s keys, $A$'s keys a strict superset of $B$'s keys,
and all other cases.
The columns are different cases of whether the action of $\otimes$ on $0_A$ and $0_B$ 
satisfies the two above constraints.
The entries indicate an upper bound on the support of $A \sjoino B$.
When the zero product property ($\forall a, b; a \otimes b = 0_A \otimes 0_B \Rightarrow a = 0_A \lor b = 0_B$)
holds, they upper bounds are also lower bounds.

\begin{table}[htb]
\centering
\begin{tabular}{c|cccc}
$\forall a; a \otimes 0_B = 0_A \otimes 0_B$ & False & True  & False & True \\
$\forall b; 0_A \otimes b = 0_A \otimes 0_B$ & False & False & True  & True \\
\hline
$K_A = K_B$                & $\subseteq S_A \union S_B$ & $\subseteq S_B$ & $\subseteq S_A$ & $\subseteq S_A \cap S_B$ \\
$K_A \subsetneq K_B$          & Unbounded & $\subseteq S_B$ & Unbounded & $\subseteq S_A \cap S_B$ \\
$K_A \supsetneq K_B$          & Unbounded & Unbounded & $\subseteq S_A$ & $\subseteq S_A \cap S_B$ \\ 
Otherwise & Unbounded & Unbounded & Unbounded & $\subseteq S_A \times S_B$ \\
\hline 
\multicolumn{1}{c}{} & \multicolumn{4}{c}{(Replace $\subseteq$ with $=$ if zero product property holds)} \\ 
\end{tabular}
\caption{Upper bounds on $\supp(A \sjoino B)$. $S_A = \supp(A)$. $S_B = \supp(B)$.}
\label{tZeroConstraints}
\end{table}

We conclude that we may loosen the constraints on $\otimes$ 
when we know one of the key conditions between $A$ and $B$ hold
that guarantee us bounded support.
This enables us to express the CombBLAS Scale operation on sparse vectors
as a join between a matrix with default value 0 and a vector with default value 1.

\clearpage
\section{Algorithms written in Lara}
\subsection{Markov Chain Clustering}

\begin{align*}
&\text{\textbf{Input}: matrix $matA$ with schema: row col | value} \\
&\text{\textbf{Constants}: } prunelimit, epsilon \\
&oldchaos := newchaos := 1000 \\
&\textbf{do} \text{ \{} \\
&\quad \!\begin{aligned}[t]
&oldchaos := newchaos \\
&AxA := \rho_{col' \to col} ((matA \sjoin_* \rho_{col \to col', row \to col}(matA) ) \union_+ E_{row,col'}) \\
&squareA := \map_{value := value^2}(AxA) \\
&colsums := squareA \union_+ E_{col} \\
&tempA := squareA \sjoin_{/_0} colsums \quad \text{ where } a \;/_0\; b := \bif b = 0 \bthen 0 \belse a / b \\
&prunedA := \map_{value := \bif value > prunelimit \bthen value \belse 0}(tempA) \\
&colssqs := \map_{sumSquare := value^2}(prunedA) \union_+ E_{col} \\
&colmaxs := \map_{maxVal := value} (prunedA \union_{\max} E_{col}) \\
&newchaos := \map_{value := maxVal - sumSquare} (colmaxs \join colssqs) \union_{\max} E \\
&matA := prunedA \\
\end{aligned} \\
&\text{\} } \textbf{while } (oldchaos - newchaos > epsilon) \\
&\textbf{Output } matA 
\end{align*}

\subsection{LU Decomposition}
Here is an LU Decomposition algorithm without pivoting. 
This version assumes the input matrix $A$'s diagonal is all nonzero.
\begin{align*}
&\text{\textbf{Input}: $N \times N$ matrix $A$ with schema: r c | v} \\
&L := Ident(N) \\
&\textbf{for } j:=1 \text{ to } N-1  \text{ \{} \\
&\quad \!\begin{aligned}[t]   
  &T := Ident(N) \\
  &\textbf{parfor } i:=j+1 \text{ to } N  \text{ \{} \\
  &\quad \!\begin{aligned}[t]
    &r := A(i,j) / A(j,j) \\
    &L(i,j) := r \\
    &T(i,j) := -r \\
  \end{aligned} \\
  &\text{\}} \\
  &A := T \;{+.*}\; A \\
\end{aligned} \\
&\text{\} } \\
&U := A \\
&\textbf{Output } L, U 
\end{align*}

The $Ident$ call refers to an identity matrix of size $N$.
There is no representation for creating this matrix without a constructor.

The division $A(i,j) / A(j,j)$ acts on scalars.  
I think this works with a zero-dimensional array using $\sjoin_/$.

The assignment $L(i,j) := r$ is a $\map$ 
with a function that updates $v$ only for keys $i$ and $j$, 
leaving the other values untouched.

The matrix multiply $T \;{+.*}\; A$ step is a macro for 
$\rho_{c' \to c} ((T \sjoin_* \rho_{c \to c', r \to c}(A)) \union_+ E_{r,c'})$.



\subsection{(incomplete) User-History PageRank from a Web Crawler}
Suppose we have a data structure from a web crawler 
that stores the website content for many websites, 
as in Table~\ref{fCrawler}.
Also suppose we have Table~\ref{fHistory} storing the Unix timestamp
that a user (determined by user identifier) has last visited each site.
Our goal is to 
\begin{enumerate}
\item Construct an adjacency matrix (Table~\ref{fLink}) from the web crawler table,
where each row contains the sites that a particular site links to.
We model this as a black-box function \emph{parse},
which takes the content of a webpage and outputs a set of links 
the webpage links to.
\item For a particular user, 
determine the Personalized PageRank of sites weighted by 
the 10 sites a user visited most recently.
\end{enumerate}
The Personalized PageRank is the same as normal PageRank 
except that the ``random restarts'' go to the sites in the 
user's browsing history, weighted by how recently they have been visited.
See Section 1.1.1 of 
\url{http://research.microsoft.com/pubs/145447/mod113-bahmani.pdf}.

\begin{figure}[htb]
\centering
\subfloat[C]{ 
  \begin{tabular}{c|c}
   & [`'] \\
  site & content \\
  \hline
  a.com & <html>...</html> \\
  b.org & <html>...</html> \\
  \vdots & \vdots \\
  \end{tabular}
  \label{fCrawler}
}
\subfloat[H]{ 
  \begin{tabular}{cc|c}
   & & [0] \\
  pid & site & lastVisit \\
  \hline
  p01 & b.org & 1450922135 \\
  p01 & c.com & 1447829664 \\
  \vdots & \vdots & \vdots \\
  \end{tabular}
  \label{fHistory}
}
\subfloat[L]{ 
  \begin{tabular}{cc|c}
   & & [0] \\
  site & link & v \\
  \hline
  a.com & b.org & 1 \\
  a.com & c.com & 1 \\
  c.com & f.com & 1 \\
  \vdots & \vdots & \vdots \\
  \end{tabular}
  \label{fLink}
}
\caption{Tables in the Personalized PageRank workflow}
\label{fPersonalPageRankBuild}
\end{figure}

\end{document}